%% file: layer-extraction-arxiv.tex
\documentclass{egpubl}
\usepackage{eg2015}

 \ConferencePaper        
 \electronicVersion 

\ifpdf \usepackage[pdftex]{graphicx} \pdfcompresslevel=9
\else \usepackage[dvips]{graphicx} \fi

\PrintedOrElectronic

\usepackage{t1enc,dfadobe}

\usepackage{egweblnk}
\usepackage{cite}
\usepackage{amsmath}
\usepackage[export]{adjustbox}
\usepackage{float}
 \usepackage{multirow,bigstrut}

\title[LayerBuilder: Layer Decomposition for Interactive Image and Video Color Editing]{LayerBuilder: Layer Decomposition for \\ Interactive Image and Video Color Editing}

\author[S. Lin et al.]
       {
        Sharon Lin$^1$$^2$ Matthew Fisher$^1$$^3$ Angela Dai$^1$ Pat Hanrahan$^1$
        \\
         $^1$Stanford University\\
         $^2$Zazzle\\
         $^3$Adobe Research\\
       }

\begin{document}


\vspace{-2em}

\teaser{
\vspace{-2em}
    \begin{tabular}{cccc}
    \includegraphics[width=.2\linewidth]{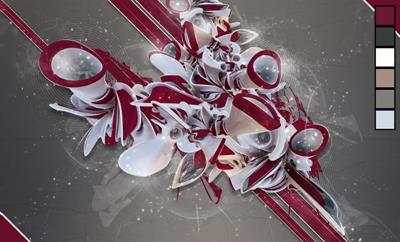} & \includegraphics[width=.31\linewidth]{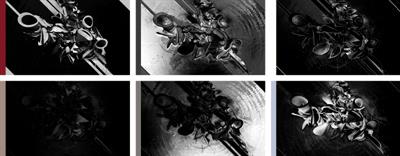} & \includegraphics[width=.2\linewidth]{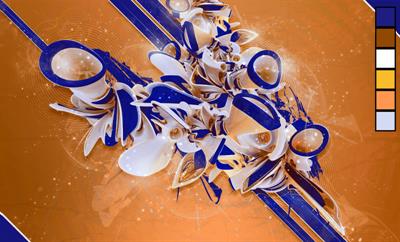} & \includegraphics[width=.2\linewidth]{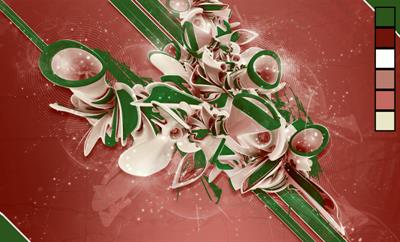} \\
    \includegraphics[width=.2\linewidth]{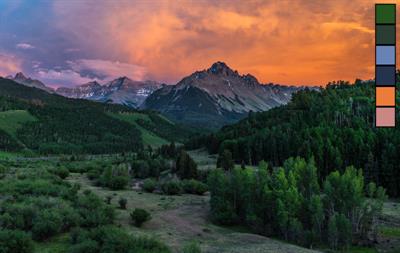} & \includegraphics[width=.307\linewidth]{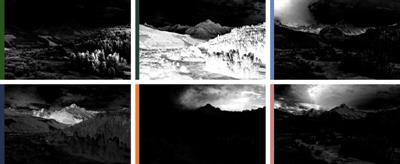} & \includegraphics[width=.2\linewidth]{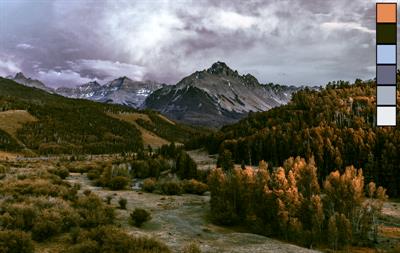} & \includegraphics[width=.2\linewidth]{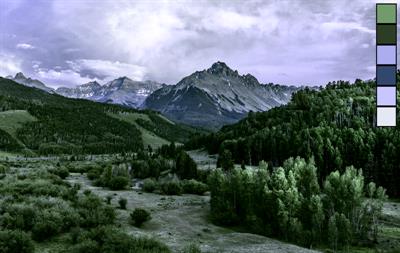} \\
    Input Image & Layer Decomposition &  \multicolumn{2}{c}{Recolorings} \\
    \end{tabular}

   \caption{LayerBuilder takes an input image and automatically computes a set of colored layers whose linear combination reconstructs the original image. Expressed in this form, it is straightforward to interactively recolor the image by choosing a new set of colors for the layers. On the right, we show two recolorings of each input image. The layer colors are shown in the upper-right of each image. Layer weights range from 0 (black) to 1 (white).}
  \label{fig:teaser}
 }

\maketitle

\begin{abstract}
Exploring and editing colors in images is a common task in graphic design and photography. However, allowing for interactive recoloring while preserving smooth color blends in the image remains a challenging problem. We present LayerBuilder, an algorithm that decomposes an image or video into a linear combination of colored layers to facilitate color-editing applications. These layers provide an interactive and intuitive means for manipulating individual colors. Our approach reduces color layer extraction to a fast iterative linear system. Layer Builder uses locally linear embedding, which represents pixels as linear combinations of their neighbors, to reduce the number of variables in the linear solve and extract layers that can better preserve color blending effects. We demonstrate our algorithm on recoloring a variety of images and videos, and show its overall effectiveness in recoloring quality and time complexity compared to previous approaches. We also show how this representation can benefit other applications, such as automatic recoloring suggestion, texture synthesis, and color-based filtering.

\begin{classification}
  \CCScat{I.4.9}{Image Processing and Computer Vision}{Applications};
\end{classification}

\end{abstract}


\newcommand*{\showBigFigures}{}
\newcommand{\remark}[1]{}
\newcommand{\owner}[1]{{\color{blue}[#1]}}
\newcommand{\note}[1]{{\color{red}[#1]}}

\newcommand*\rot{\rotatebox{90}}
\def\imagetop#1{\vtop{\null\hbox{#1}}}

\input{introduction.tex}

\input{background.tex}

\input{algorithm.tex}

\input{results.tex}

\input{applications.tex}

\input{discussion.tex}

\bibliographystyle{eg-alpha-doi}
\bibliography{layer-extraction}

\end{document}

%% file: introduction.tex
\section{Introduction}

Representing an image by separating different elements into layers is an integral feature of digital image editing software such as Adobe Photoshop or GIMP. Layers offer an intuitive way to edit the color and geometry of independent components and localize changes to the desired portion of the final image. When creating a digital image, artists carefully maintain the set of layers to make it easy to perform edits that will enable rapid iteration through the design space. In most cases, combining the layers is computationally straightforward, allowing an artist to interactively evaluate how layer edits affect the final image.

A layer representation is a powerful editing tool, but it is not always available. Once an image is rasterized, its decomposition into layers is typically lost. Furthermore, photographs or videos captured by camera or produced by rendering systems may not have an accessible decomposition into meaningful layers. The problem of recovering a good set of layers for an image or video is not easy to specify, as different users and applications will prefer different layerings. Nevertheless, for flattened images and videos, we would still like to recover some of the powerful interactive editing operations enabled by a meaningful layer decomposition.

In this paper, we focus on the problem of extracting layers for use in image and video color editing. Our system takes an input image or video and computes a decomposition into colored layers, whose combination reconstructs the original input. Users can optionally specify additional constraints to further refine the layers. Given this representation, it is easy to perform edits and interactively recompose the final recolored result. Figure~\ref{fig:teaser} visualizes the process for both photographic and artistic images. 

Our layers enable easy and interactive editing of images and videos, even in the presence of challenging color blending effects such as semi-transparency, motion, and unfocused pixel blur. LayerBuilder enables finer control over each color range compared to color transfer techniques~\cite{Reinhard2001,Tai2005local}, which require a reference image, and material or object matting techniques~\cite{KNNMatting}, which tend to extract larger patterned regions. Recoloring methods using affinity-based edit propagation~\cite{AppProp,InstantProp,PalettePhotoRecoloring} have similar control over colors but produce more halo artifacts in color blending regions because they do not necessarily preserve color interpolation relationships. Manifold-preserving edit propagation reduces these artifacts but requires solving a linear system for each color change~\shortcite{MPEP}. Our method handles these regions as well as manifold-preserving edit propagation, while making color editing interactive after layer setup. 

We show how to use the concept of locally linear embedding to reduce the problem of extracting layers for color editing to a simple iterative linear system. We adapt and accelerate the edit propagation technique from Chen and colleages~\shortcite{MPEP} to work for computing layers. Instead of performing our solve directly over pixels, we demonstrate that it is sufficient to solve for the layer decomposition on a much smaller set of superpixels, then extrapolate the pixel decomposition from the superpixel solution. For the images shown, this results in a system with three orders of magnitude fewer free variables, permitting solves on large images with many layers while remaining within reasonable memory and computational budgets. Finally, we compare our method against previous recoloring approaches to demonstrate its interactivity and robustness to color blending and parameter choice. As shown in the supplemental video, once the layer decomposition has been performed, all video and image color edits are interactive, allowing artists to quickly explore the potential coloring space.

We demonstrate the effectiveness of our system for recoloring a variety of artistic and natural images. We further show that our layers can be used to augment existing image processing or synthesis applications. We use LayerBuilder to extend work on automatic pattern coloring suggestion, which previously required pre-layered images and operated on an unblended set of colors, to support any input image and take advantage of our continuous layer decomposition~\cite{PatternColoring}. Texture synthesis algorithms which operate on a general feature space can easily use the layer decomposition as a local feature vector for each pixel~\cite{TextureSynthesis}, allowing synthesized textures to more faithfully respect long-range effects in the exemplar. Finally, we show that like with digital image editing tools, it is easy to produce significant but localized modifications to the image by directly filtering the layers and recombining them.

%% file: background.tex
\section{Related Work and Background}

Related work for our method spans two main areas: image recoloring via edit propagation and image layerization. We also provide some background on locally linear embedding.

\subsection{Image Recoloring via Edit Propagation}

Affinity-based recoloring methods use pixel similarity to propagate user color edits, in the form of strokes or palette changes, to the rest of the image. Levin and colleagues consider the similarity between neighboring pixels based on luminance~\shortcite{ScribbleColorization}. To handle textured regions, An and Pellacini consider the similarity between all pairs of pixels~\shortcite{AppProp}, a computation that can be sped up by clustering pixels beforehand~\cite{KDTreeProp}. Some methods compute an alpha influence matte for each stroke ~\cite{ScribbleBoost,DiffusionMaps}, allowing quick subsequent color tweaks. Li and colleagues achieve interactive recoloring time by formulating edit propagation as a radial basis function (RBF) interpolation problem where function coefficients depend on user-edited pixels~\shortcite{InstantProp}. Chang and colleagues focus on recoloring via extracting and editing the image's color palette~\shortcite{PalettePhotoRecoloring}. They introduce a faster recoloring method by using RBF functions where the coefficients depend only on the input palette colors, and spatial information is not considered. However, recoloring by shifting pixel colors based on their similarity to other pixels (or palette colors) does not necessarily preserve the color interpolation relationships between pixels along color transitions (e.g. edges and gradients). As a result, these methods tend to produce halo artifacts in these color blending regions~\cite{MPEP}. 

To reduce halo effects, Chen and colleagues propose a smoothness term that preserves the pixel manifold structure of the image~\shortcite{MPEP}. Each pixel's color is represented by a weighted combination of its $K$-nearest pixels in appearance and space. Recoloring an image should then both respect user input as well as maintain these color relationships in the result. This method can be accelerated by solving edit propagation for a smaller set of representative samples and then interpolating samples to get the final pixel colors~\cite{MPEPSparse}. Our work takes this idea of preserving the pixel manifold up one level of indirection when decomposing an image into layers. The layer decomposition of a pixel should be a weighted combination of its $K$ neighbors' layer decomposition, and the weights should be determined by the color relationships of the pixels. Once we compute the layers, image recolorization is very fast, and only requires adding layers together. 

\subsection{Image Layers}

A similar notion of layers is used in image matting~\cite{BayesianMatting,ClosedFormImageMatting,MultipleImageLayers}, which separates objects from the background, and material matting~\cite{MaterialMatting,KNNMatting}, which separates different textures. A stroke-based or trimap input is used to estimate alpha mattes for each layer as well as layer colors. These techniques focus on facilitating object or material manipulation, and their layers tend to capture objects or textures that may have multiple colors. Our layers represent a single color, allowing for cleaner and more fine-grained color changes as we demonstrate in Section~\ref{sec:comparison}. Both types of layers have orthogonal goals and may be combined to better localize color edits to similarly-colored objects or materials.

Carroll and colleagues aim to recolor objects in photographs while updating their interreflections with other objects consistently~\shortcite{Carroll2011illumination}. Their approach involves first breaking the photo down into its material reflectance and illumination intrinsic images~\cite{Bousseau2009User}. Assuming the material reflectance image is piecewise constant and thus easily recolorable, the authors focus on decomposing the illumination image into color-based layers given a user-specified set of main colors. This method works best for recoloring objects that are in focus where color blending is explained mainly by illumination. We use a similar, but simpler, energy function for computing color-based layers. Rather than specializing on interreflections, our color-based layers target a more general class of color blending effects, such as ones due to depth-of-field, motion blur, and including more pronounced interreflections. Our color-based layers can create plausible results for a wide range of color edits without requiring a decomposition into reflectance and illumination images. In Section~\ref{sec:comparison}, we compare our method to the one by Carroll and colleages for the case of interreflections. 

Concurrently, Tan and colleagues developed a method for extracting ordered single-color layers that use the more traditional over operator to reconstruct the original image~\shortcite{TAN2016DIL}. In contrast, our method extracts order-agnostic additive layers that, although do not support occlusion-related edits, involve minimizing a simpler energy function and still enable a variety of color edits.

\subsection{Locally Linear Embedding}
\label{sec:LLE}

Locally linear embedding (LLE) is a non-linear dimensionality reduction technique for data that lie on a manifold embedded in a higher-dimensional space. LLE is based on the intuition that each point and its neighbors form a locally linear patch of the manifold~\cite{LLE}. Then for a set of data points $x_1,...,x_n$, we can approximate each $x_i$ as a linear combination of its $K$ nearest neighbors. Specifically, we compute the weights $w_{ij}$ that minimize the residual sum of squares from reconstructing each $x_i$,
\[ \sum_{i=1}^n \big\lVert x_i - \sum_{j\neq i}w_{ij}x_j \big\rVert^2 \]
where for each $i$, $\sum_j w_{ij} = 1$.
These weights characterize the manifold structure and can be computed efficiently~\cite{LLE}. This idea has been used to propagate scribble edits through images and video, noting that in natural images, the color at a pixel is often a blend of the colors of neighboring pixels~\cite{MPEP}.

%% file: algorithm.tex
\section{Layer Decomposition}

\newcommand{\layerCount}{\mathsf{N}}
\newcommand{\pixelCount}{\mathsf{P}}
\newcommand{\superpixelCount}{\mathsf{S}}
\newcommand{\pixel}{p}
\newcommand{\superpixel}{s}
\newcommand{\pixelNeighborCount}{\mathsf{K}_p}
\newcommand{\superpixelNeighborCount}{\mathsf{K}_s}
\newcommand{\superpixelLayerWeights}{L}
\newcommand{\pixelLayerWeights}{X}
\newcommand{\video}{\mathbb{V}}
\newcommand{\pixelManifold}{Q}
\newcommand{\pixelManifoldElem}{q}
\newcommand{\superpixelManifoldElem}{w}

\newcommand{\manifoldTermMatrix}{M}
\newcommand{\superpixelManifold}{W}
\newcommand{\reconstructionTermMatrix}{R}
\newcommand{\layerColors}{c}
\newcommand{\superpixelColors}{B} 
\newcommand{\sumOneTermMatrix}{U}
\newcommand{\sumOneTermMatrixElem}{u}
\newcommand{\userTermMatrix}{E}
\newcommand{\userTermTarget}{T}
\newcommand{\userConstraintCount}{\mathsf{C}} 

Our goal is to take an input image or video representing a set of $\pixelCount$ pixels, a target number of layers $\layerCount$, and compute the following:

\begin{itemize}
  \item A set of $\layerCount$ layer colors
  \item For each of the $\layerCount$ layers, a $\pixelCount \times 1$ column vector $X_j$ whose entries denote the contribution of layer $j$ to each pixel. $X$ is the $\pixelCount \layerCount \times 1$ column vector formed by concatenating all $X_j$.
\end{itemize}

\begin{figure}[tb]
\ifdefined\showBigFigures
\renewcommand{\tabcolsep}{3pt}
\begin{tabular}{cc}
	\includegraphics[width=.45\linewidth]{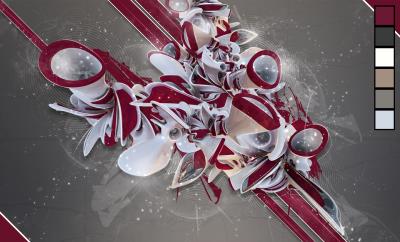} &
	\includegraphics[width=.45\linewidth]{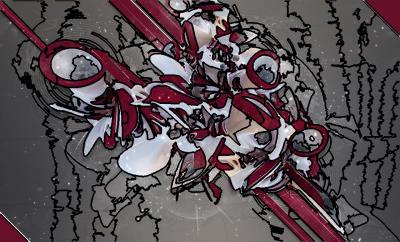} \\
	(a) Input image \vspace{1ex} & (b) Superpixel segmentation \\
	\includegraphics[width=.45\linewidth]{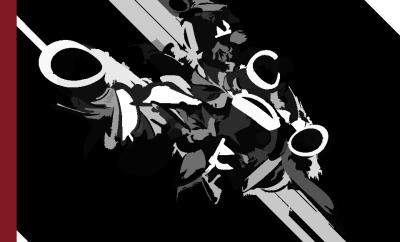} &
	\includegraphics[width=.45\linewidth]{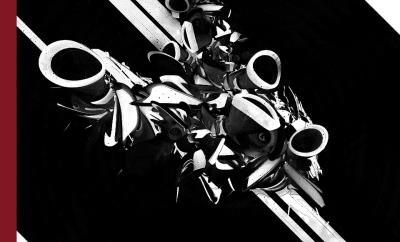} \\
	(c) Per-superpixel layer weights & (d) Per-pixel layer weights \\
\end{tabular}
\fi
\caption{Algorithm overview. (a) the input image and the extracted palette. (b) the superpixel segmentation using 250 superpixels. (c) the red layer weights for each superpixel, computed using Equation~\ref{eq:finalSolve}. (d) the result shown in (c) is used to compute per-pixel weights for the red layer using Equation~\ref{eq:pixelProj}. Figure~\ref{fig:teaser} visualizes the complete layer decomposition for this image using 2000 superpixels.}
\label{fig:steps}
\vspace{-2em}
\end{figure}

Our algorithm is summarized in Figure~\ref{fig:steps} and consists of four main steps. First, compute the layer colors from the input pixels or obtain them from the user (Figure~\ref{fig:steps}a). Second, compute a superpixel segmentation of the image~(Figure~\ref{fig:steps}b). Third, solve for the layer weights for each superpixel. This problem reduces to solving a linear system (Figure~\ref{fig:steps}c). Finally, for each layer, compute a per-pixel weight by linearly combining the set of per-superpixel weights (Figure~\ref{fig:steps}d). 

\subsection{Layer Colors}
\label{sec:color}

Because people may have different goals when recoloring an image, we allow the user to either specify a palette or use an automatic one. Automatic palettes provide a useful starting point and also enable automatic recolorings (Section~\ref{sec:patternRecoloring}). In all results, we specify when an automatic palette is used.

To provide intuitive controls over the space of possible colorings, our algorithm should choose the $\layerCount$ colors that a person would likely choose as well as choose colors that can linearly reconstruct the original image in RGB (or LAB) space. To this end, we adapt the color palette model introduced by Lin and Hanrahan, which has been trained to match palettes people would extract from images~\shortcite{SharonPaletteExtraction}. We add a penalty term to this model based on the average color reconstruction error, where error per pixel is measured by the distance from the pixel color to the convex hull of the palette. In our experiments, we weight the term by $-0.01\lambda$ where $\lambda$ is the sum of absolute weights in the Lin and Hanrahan model.

\subsection{Superpixel Segmentation}
\label{sec:superpixel}

Computing the per-layer weights directly on pixel values results in a system with a free variable for each pixel and each layer weight. This rapidly becomes intractable to solve directly: for large images this may result in millions of free variables, growing to billions of free variables for video sequences. To combat this problem, we take advantage of the high correlation between nearby pixels. We solve for the layer weights on a much smaller number of superpixels (or supervoxels), and then compute the layer solution for each pixel using a linear combination of nearby superpixels. The image and video results in this paper use at most 4000 superpixels, resulting in systems with multiple orders of magnitude fewer free variables.

To compute superpixels, we use a variation of the seeded region growing algorithm (SRG)~\cite{SRG} with iterative $k$-means re-centering. We manually specify the number of superpixels $\superpixelCount$ and initialize each to a random seed location. Superpixels are then computed by SRG, and their centroids and colors are used as seeds for the next iteration of SRG, repeating for a total of 5 $k$-means iterations. For supervoxels, we define pixel neighbors through 6-connectivity: 4 spatial and 2 temporal neighbors.

From each superpixel we compute a 6-dimensional feature vector (r,g,b,x,y,t) formed by concatenating the average pixel color with the superpixel's spatiotemporal centroid. The color, spatial, and temporal coordinates are normalized between 0 and 1 and re-weighted according to importance. In our experiments, we weight the spatial coordinates (x,y) by 0.5, to bias distance calculations towards color and temporal similarity. Figure~\ref{fig:steps}b shows an example superpixel segmentation with superpixel boundaries outlined in black.

\subsection{Energy Function}
\label{sec:energy}

In this section we will compute a set of per-superpixel layer contributions $\superpixelLayerWeights_j \in \superpixelLayerWeights$, where $L_j$ is a $\superpixelCount \times 1$ column vector denoting the contribution of layer $j$ to each superpixel, and $\superpixelLayerWeights$ is the $(\superpixelCount \layerCount) \times 1$ column vector formed by concatenating all $L_j$. To quantitatively define what constitutes a good layering, we minimize an energy function $\Theta(L)$ over the superpixel-layer contributions $L$. $\Theta(L)$ is a weighted sum of four terms: manifold consistency $M(L)$, image reconstruction $R(L)$, unity $U(L)$, and explicit constraints $E(L)$.
\begin{equation}
\Theta(L) = \lambda_{m} M(L) + \lambda_{r} R(L) + \lambda_{u} U(L) + \lambda_{e} E(L)
\end{equation}

\textbf{Manifold consistency}. Layer contributions should be locally consistent, in that nearby and similar superpixels should have similar layer decompositions. We implement this term using the superpixel manifold determined by locally linear embedding, similar to the per-pixel constraints used for edit propogation~\cite{MPEP}.

Following the approach used in Section~\ref{sec:LLE}, we express the color of each superpixel $\superpixel_i$ as a linear combination of the color of the $\superpixelNeighborCount$-nearest superpixels $\superpixel_{ij}$:
\begin{equation*}
 \bigg\lVert \text{Color}(\superpixel_{i}) - \sum_{j=1}^{\superpixelNeighborCount} \superpixelManifoldElem_{ij} \text{Color}(\superpixel_{ij}) \bigg\rVert^2
\end{equation*}
We solve for the weights $\superpixelManifoldElem_{ij}$ and represent them in an $\superpixelCount \times \superpixelCount$ superpixel-manifold matrix $\superpixelManifold$.
The manifold consistency term is then:
\begin{equation}
\begin{array}{lcl}
M(\superpixelLayerWeights) & = & \sum\limits_{j=1}^{\layerCount}||(I_\superpixelCount-\superpixelManifold)\superpixelLayerWeights_j||^2 \\
	 &&\\
     & = & ||(I_{\superpixelCount\layerCount} - \manifoldTermMatrix)\superpixelLayerWeights||^2
\end{array}
\end{equation}
where $I_n$ is the $n \times n$ identity matrix and $\manifoldTermMatrix$ is a block diagonal matrix with overall dimension $\superpixelCount\layerCount \times \superpixelCount\layerCount$ of the form:
\begin{equation}
\manifoldTermMatrix = \left( 
	\begin{array}{cccc} 
	\superpixelManifold & 0 & \hdots & 0 \\
	0 & \superpixelManifold & \hdots & 0 \\
	\vdots & \vdots & \ddots &\vdots \\
	0 & 0 & \hdots & \superpixelManifold 
	\end{array}
\right )
\end{equation}

\textbf{Image reconstruction}. The layers should reconstruct the original image when multiplied by their colors and added together. We enforce this by measuring the squared image reconstruction error:
\begin{equation}
\begin{array}{lcl}
R(\superpixelLayerWeights) & = & \sum\limits_{d \in (r,g,b)}||\sum\limits_{j=1}^{\layerCount}(\layerColors_{dj}\superpixelLayerWeights_j) - \superpixelColors_d||^2 \\
	 &&\\
     & = & ||\reconstructionTermMatrix\superpixelLayerWeights - \superpixelColors||^2
\end{array}
\label{eq:recon}
\end{equation}
where $\superpixelColors_d$ is the $\superpixelCount \times 1$ vector of channel $d$ superpixel colors and $c_{dj}$ is channel $d$ of the color of layer $j$. $\reconstructionTermMatrix$ is a block matrix with overall dimension $3\superpixelCount \times (\superpixelCount\layerCount)$:
\begin{equation}
\reconstructionTermMatrix = \left( 
	\begin{array}{cccc} 
	\layerColors_{r1}I_{\superpixelCount} & \layerColors_{r2}I_{\superpixelCount} & \hdots & \layerColors_{r\layerCount}I_{\superpixelCount} \\
	\layerColors_{g1}I_{\superpixelCount} & \layerColors_{g2}I_{\superpixelCount} & \hdots & \layerColors_{g\layerCount}I_{\superpixelCount} \\
	\layerColors_{b1}I_{\superpixelCount} & \layerColors_{b2}I_{\superpixelCount} & \hdots & \layerColors_{b\layerCount}I_{\superpixelCount}
	\end{array}
\right )
\end{equation}
and $\superpixelColors$ is the concatenation of all $\superpixelColors_d$.

\textbf{Unity}. We encourage layer contributions for each superpixel to sum up to one, to help normalize the total contribution that can affect a single superpixel. Formally:
\begin{equation}
U(L) = ||\sumOneTermMatrix\superpixelLayerWeights - \hat{1}||^2
\label{eq:unity}
\end{equation}
where $\hat{1}$ is a $\superpixelCount \times 1$ column vector of ones and $\sumOneTermMatrix$ is a $\superpixelCount \times (\superpixelCount\layerCount)$ indicator matrix whose rows correspond to superpixels, with ones in each column index corresponding to that superpixel's layer contributions:
\[ \sumOneTermMatrixElem_{ij} = \begin{cases} 1 &\mbox{if } j - i \equiv 0 \pmod{\superpixelCount} \\ 
0 & \mbox{if } j - i \not\equiv 0 \pmod{\superpixelCount} \end{cases} \]

\textbf{Explicit constraints}. In some cases, hints from the user indicating that a given layer should contribute to a given image region can help create a better layering. For example, these hints can separate similarly-colored but semantically different regions. Given $\userConstraintCount$ user constraints, we penalize layer contributions when they deviate from user constraints:
\begin{equation}
E(L) = ||\userTermMatrix\superpixelLayerWeights - \userTermTarget||^2
\label{eq:explicit}
\end{equation}
where $\userTermMatrix$ is a $\userConstraintCount \times (\superpixelCount\layerCount)$ indicator matrix that selects the user-specified superpixel-layer contributions from $\superpixelLayerWeights$, and $\userTermTarget$ is a length $\userConstraintCount$ vector containing the user-specified target values.

\begin{figure*}[tb]

\ifdefined\showBigFigures
\renewcommand{\tabcolsep}{4.5pt}
  \begin{tabular}{c|cccccc}
  \centering
  \multirow{3}{*}{\imagetop{\includegraphics[width=.283\linewidth]{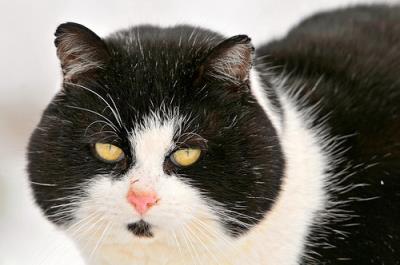}}}&
  &
  \imagetop{\includegraphics[width=.11\linewidth,height=0.1in]{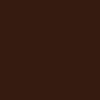}}&
  \imagetop{\includegraphics[width=.11\linewidth,height=0.1in]{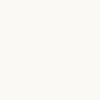}}&
  \imagetop{\includegraphics[width=.11\linewidth,height=0.1in]{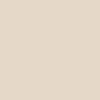}}&
  \imagetop{\includegraphics[width=.11\linewidth,height=0.1in]{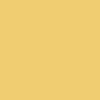}}&
  \imagetop{\includegraphics[width=.11\linewidth,height=0.1in]{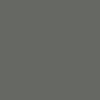}}\\
  &
  \raisebox{-3\height}{Auto}&
  \imagetop{\includegraphics[width=.11\linewidth]{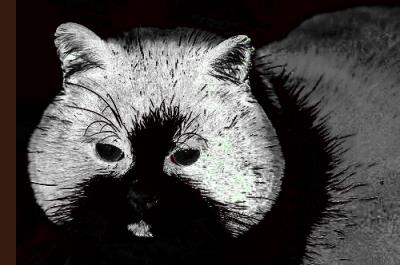}}&
  \imagetop{\includegraphics[width=.11\linewidth]{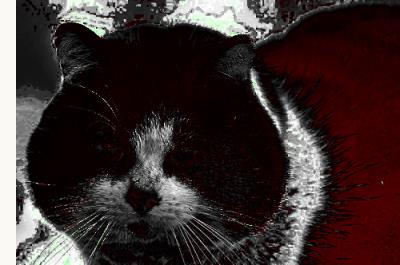}}&
  \imagetop{\includegraphics[width=.11\linewidth]{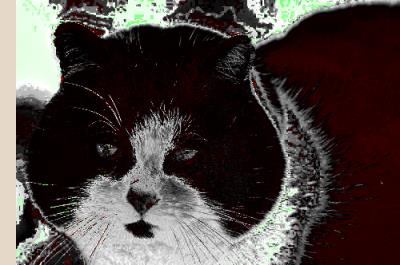}}&
  \imagetop{\includegraphics[width=.11\linewidth]{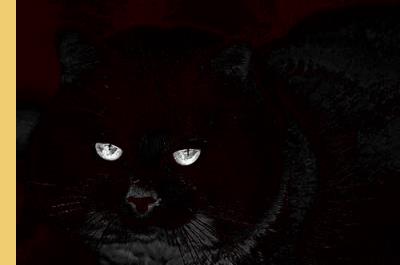}}&
  \imagetop{\includegraphics[width=.11\linewidth]{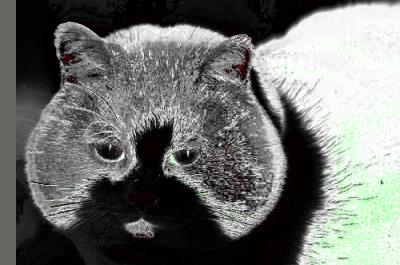}}\\
  &
  \raisebox{-3\height}{User}&
  \imagetop{\includegraphics[width=.11\linewidth]{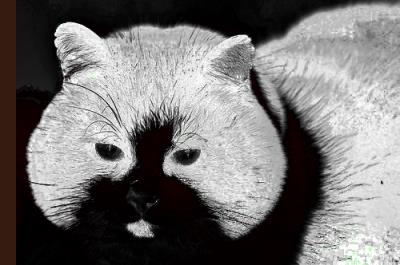}}&
  \imagetop{\includegraphics[width=.11\linewidth]{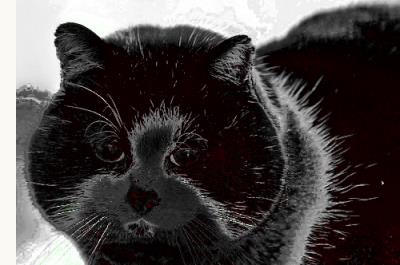}}&
  \imagetop{\includegraphics[width=.11\linewidth]{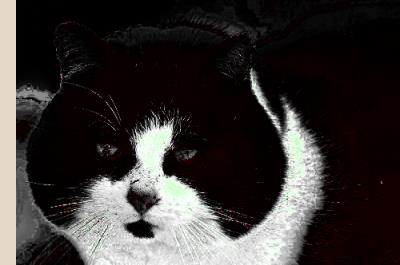}}&
  \imagetop{\includegraphics[width=.11\linewidth]{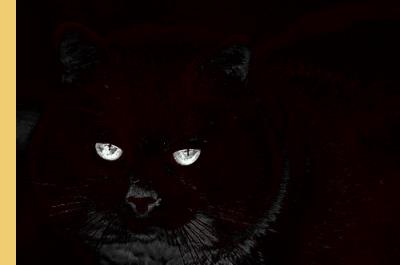}}&
  \imagetop{\includegraphics[width=.11\linewidth]{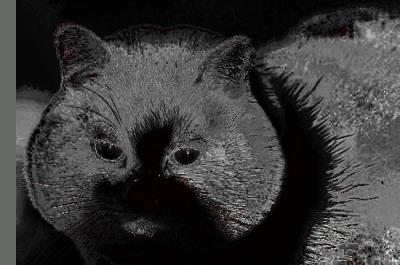}}\\
  \end{tabular}
\fi
\caption{Automatically determined constraints vs. user-specified constraints. Using user-specified constraints allows the layers to better separate the cat's white fur from the background.}
\label{fig:constraints}
\end{figure*}

When the user does not specify layer contribution constraints, our algorithm automatically adds constraints to pixels with very similar colors for each layer. This can encourage better separation between layers in many images. For trickier images with independent regions sharing similar colors, user-specified constraints are needed to better separate these regions. Figure~\ref{fig:constraints} shows a case where letting the user assign superpixels to a particular layer results in better separation between the cat's white fur and the background.

$\Theta(L)$ is a quadratic function in $L$ which can be minimized by solving a linear system:
\begin{multline}
\left( \lambda_m (I-\manifoldTermMatrix)^T(I-\manifoldTermMatrix) + \lambda_r \reconstructionTermMatrix^T\reconstructionTermMatrix + \lambda_u \sumOneTermMatrix^T\sumOneTermMatrix + \lambda_e \userTermMatrix^T\userTermMatrix \right) \superpixelLayerWeights = \\
\lambda_r \reconstructionTermMatrix^T\superpixelColors + \lambda_u \sumOneTermMatrix^T \hat{1} + \lambda_e \userTermMatrix^T\userTermTarget
\label{eq:finalSolve}
\end{multline}

Figure~\ref{fig:steps}c visualizes $\superpixelLayerWeights_j$ for one layer of the input image.

\subsection{Computing Per-pixel Weights}
\label{sec:perpixel}

We use the concept of locally linear embedding described in Section~\ref{sec:LLE} to extrapolate the per-pixel layer contributions $\pixelLayerWeights$ from the per-superpixel contributions $\superpixelLayerWeights$. We start by computing for each pixel a set of $\pixelNeighborCount$ nearest superpixel neighbors. The distance between a pixel and a superpixel is computed using the Euclidean distance between the 6-dimensional superpixel feature vector and the corresponding values for each pixel. Next, we express the color of each pixel as a linear combination of the colors of its $\pixelNeighborCount$ superpixel neighbors. Specifically, for pixel $\pixel_i$ we compute the weights $\pixelManifoldElem_{ij}$ by minimizing
\begin{equation}
 \bigg\lVert \text{Color}(\pixel_i) - \sum_{j=1}^{\pixelNeighborCount} \pixelManifoldElem_{ij} \text{Color}(\superpixel_{ij}) \bigg\rVert^2
\end{equation}
subject to the constraint $\sum_{j=1}^{\pixelNeighborCount} \pixelManifoldElem_{ij} = 1$. We represent the set of all $\pixelManifoldElem_{ij}$ with a $\pixelCount \times \superpixelCount$ matrix $\pixelManifold$. Finally, for each layer $j$ we independently compute the per-pixel layer values $\pixelLayerWeights_j$ from the per-superpixel layer values $\superpixelLayerWeights_j$ by simple matrix multiplication:
\begin{equation}
 X_j = \pixelManifold L_j
\label{eq:pixelProj}
\end{equation}

Figure~\ref{fig:steps}d shows the result of multiplying the $L_j$ visualized in Figure~\ref{fig:steps}c by $\pixelManifold$. Observe that high-frequency details such as the small slivers of red in the lower-left of the image are faithfully preserved, even though these slivers are not contained in isolated superpixels.

\begin{figure*}[htb!]
\ifdefined\showBigFigures

  \setlength{\tabcolsep}{0.5em}
	\begin{tabular}{c|cccccc}
	\multirow{4}{*}{\raisebox{-1.4\height}{\includegraphics[width=.15\linewidth]{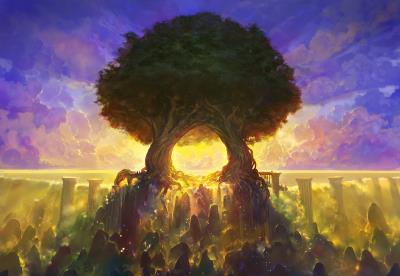}}}&
	Hard bounds&
	\includegraphics[width=.12\linewidth,height=0.1in]{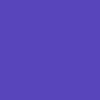}&
	\includegraphics[width=.12\linewidth,height=0.1in]{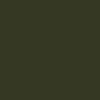}&
	\includegraphics[width=.12\linewidth,height=0.1in]{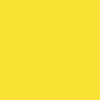}&
	\includegraphics[width=.12\linewidth,height=0.1in]{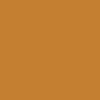}&
	\includegraphics[width=.12\linewidth,height=0.1in]{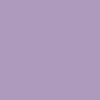} \\
	&
	\includegraphics[width=.12\linewidth]{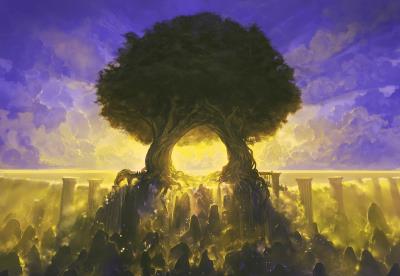}&
	\includegraphics[width=.12\linewidth]{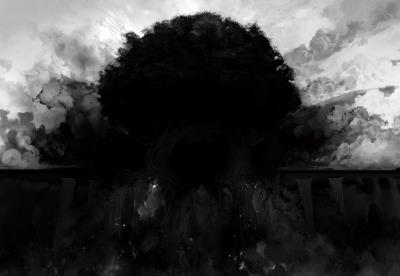}&
	\includegraphics[width=.12\linewidth]{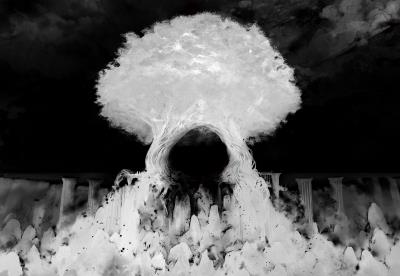}&
	\includegraphics[width=.12\linewidth]{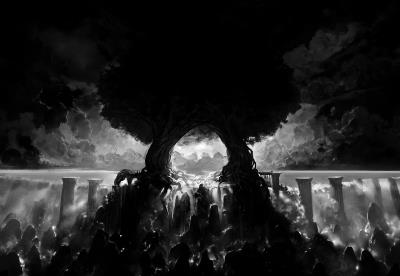}&
	\includegraphics[width=.12\linewidth]{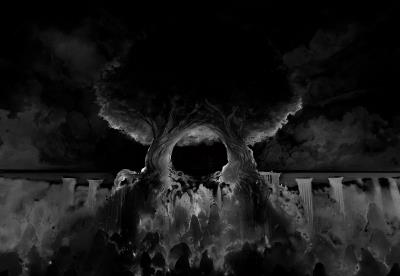}&
	\includegraphics[width=.12\linewidth]{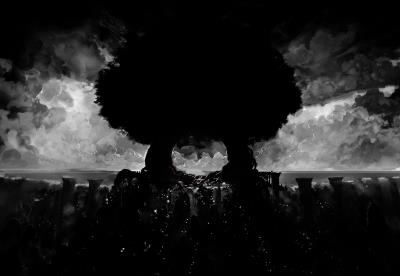} \\
	&
	Neg. Suppr. \\
	&
	\includegraphics[width=.12\linewidth]{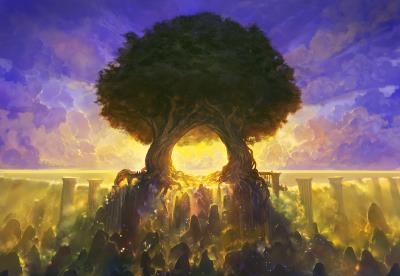}&
	\includegraphics[width=.12\linewidth]{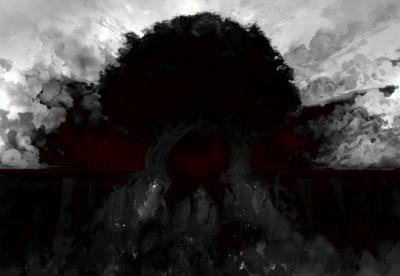}&
	\includegraphics[width=.12\linewidth]{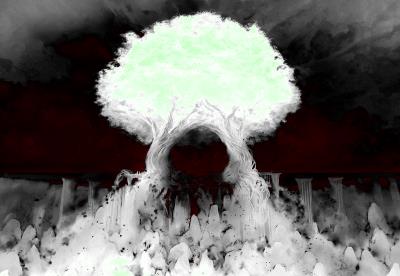}&
	\includegraphics[width=.12\linewidth]{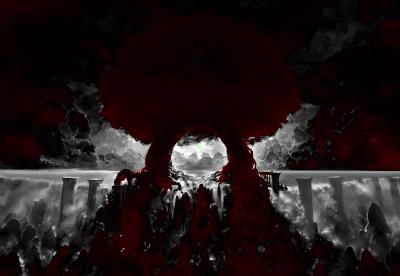}&
	\includegraphics[width=.12\linewidth]{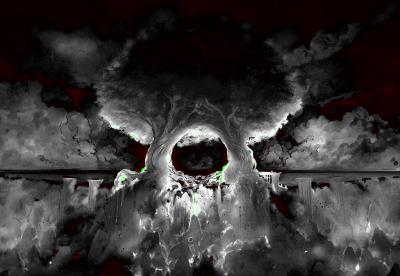}&
	\includegraphics[width=.12\linewidth]{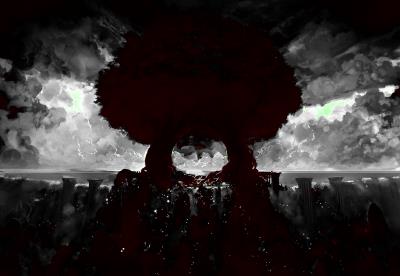} \\
	\end{tabular}
\fi

\caption{Layer and reconstruction results using 2000 superpixels. Palette colors were chosen automatically. Green indicates layer values greater than 1 while red indicates values less than 0. Using hard bounding constraints fades the orange in this image, and is more sensitive to the gamut of the palette chosen. Using soft constraints with negative suppression gives more flexibility and can better reconstruct the image. }
\label{fig:convex}

\vspace{-1em}
\end{figure*}

\subsection{Bounding Layer Contributions}

Equation~\ref{eq:finalSolve} does not constrain the range of layer contributions. Negative or large positive values can be unintuitive, causing small edits to a layer's color to have a disproportionate impact on a pixel color. One approach to solve this problem is to formulate the optimization of $L$ as a least-squares problem with hard box constraints $[0,1]$. Although problems of this form have known solutions, for large systems they can still be prohibitively slow to solve when compared against an unconstrained least-squares solve~\cite{boxConstraints}.

Another approach is to first solve the unconstrained linear problem, then add in boundary conditions that have been violated as soft constraints in future iterations of the solve. Although it requires multiple iterations, this approach can efficiently solve some types of convex problems~\cite{convexOptimization}. We first perform the unconstrained solve and check which values are negative. In the next iteration, we add a soft constraint for these values to be 0 with weight $\lambda_n$, adding to the explicit constraints in Equation~\ref{eq:explicit}. Combined with the unity term in Equation~\ref{eq:unity}, this encourages contributions to fall within the range $[0,1]$. We refer to this approach as iterative negative suppression.

We find that strictly enforcing the box constraints often makes image reconstruction from the layers very sensitive to the initial color palette. This is because the image cannot reconstruct colors outside the convex hull of the palette.
Figure~\ref{fig:convex} shows a comparison between hard box constraints (using CVX and the SDPT3 solver) and iterative negative suppression. In this example, using hard box constraints results in layers that cannot reconstruct the orange in the original image, while iterative negative suppression reconstructs the image correctly and results in an otherwise similar layering. For 4000 superpixels, the convex solver took over an hour to compute a layering, while the unconstrained linear solve took only a few seconds. All the results shown in this paper use four iterations of iterative negative suppression to bound the magnitude of layer contributions.

\subsection{Parameters}

In our experiments, we find that the weights $\lambda_m=1$, $\lambda_r=0.5$, $\lambda_u=0.1$, and $\lambda_e=0.1$, and the number of superpixel-to-superpixel and pixel-to-superpixel neighbors $\superpixelNeighborCount=30$ and $\pixelNeighborCount=10$ work well. We have not found our algorithm to be very sensitive to the choice of these parameters, although it is important that $\lambda_m$ (which preserves color blending relationships) and $\lambda_r$ (which ensures the layers reconstruct the original image) be sufficiently larger than $\lambda_u$ and $\lambda_e$. Solving for layers in RGB and LAB space also yields similar results. For the results shown, we use RGB space.

%% file: results.tex
\section{Evaluation}
\label{sec:comparison}

We evaluate our algorithm by comparing recoloring results with previous work, analyzing performance time, and showing robustness under different number of superpixels. For more comparisons, please see the supplementary materials.

\subsection{Comparisons With Edit Propagation}

\begin{figure*}[tb]
\centering
\renewcommand{\arraystretch}{0.5}
  \begin{tabular}{ccccc}
  Input & ~\cite{InstantProp} & ~\cite{MPEP} & ~\cite{PalettePhotoRecoloring} & LayerBuilder\\

  \imagetop{\includegraphics[width=.16\linewidth]{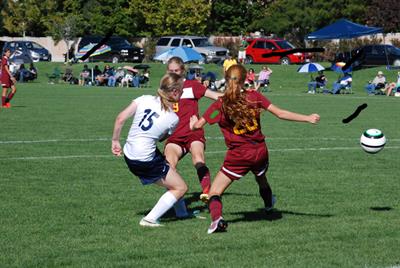}} 
  \imagetop{\includegraphics[width=.015\linewidth,frame, height=0.7in]{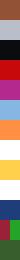}}&
  \imagetop{\includegraphics[width=.16\linewidth]{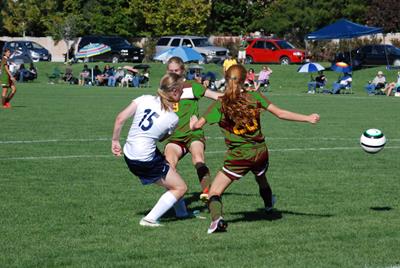}}&
  \imagetop{\includegraphics[width=.16\linewidth]{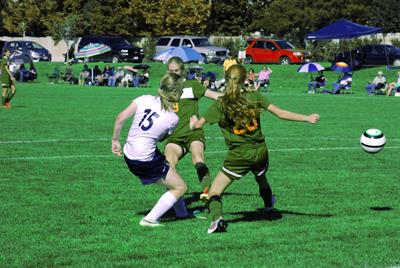}}&
  \imagetop{\includegraphics[width=.16\linewidth]{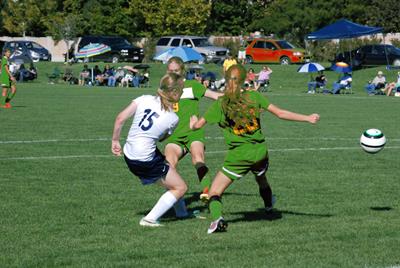}}& 
  \imagetop{\includegraphics[width=.16\linewidth]{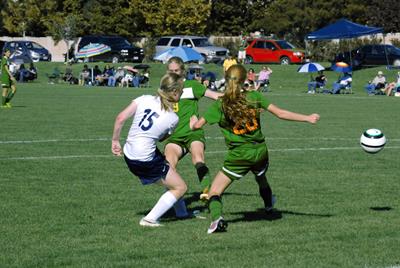}} \vspace{1pt}\\

  \imagetop{\includegraphics[width=.16\linewidth]{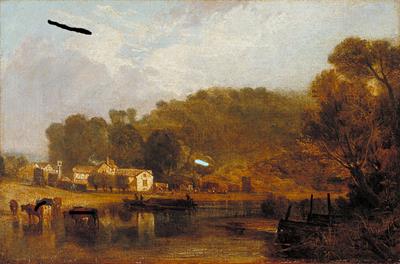}} 
  \imagetop{\includegraphics[width=.015\linewidth,frame]{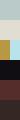}}&
  \imagetop{\includegraphics[width=.16\linewidth]{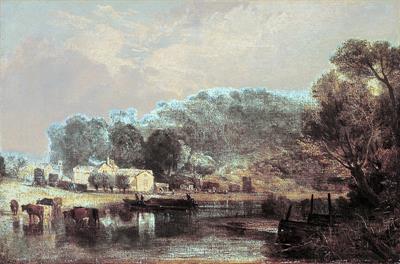}}&
  \imagetop{\includegraphics[width=.16\linewidth]{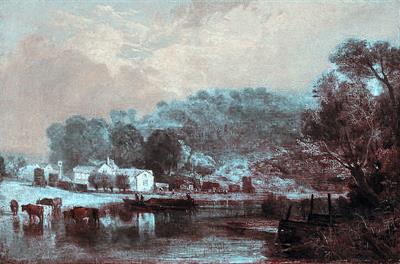}}&
  \imagetop{\includegraphics[width=.16\linewidth]{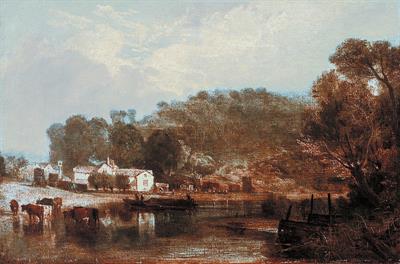}}&
  \imagetop{\includegraphics[width=.16\linewidth]{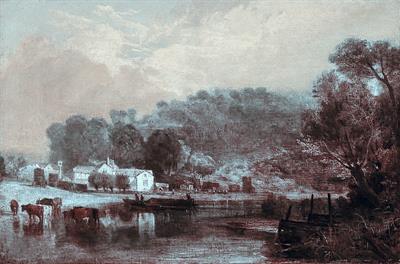}} \vspace{1pt}\\

  \imagetop{\includegraphics[width=.16\linewidth]{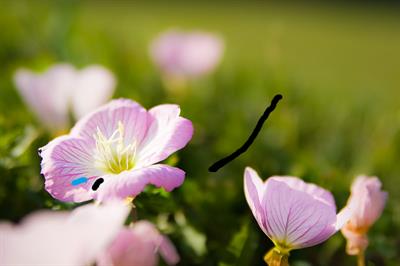}} 
  \imagetop{\includegraphics[width=.015\linewidth,frame]{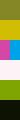}}&
  \imagetop{\includegraphics[width=.16\linewidth]{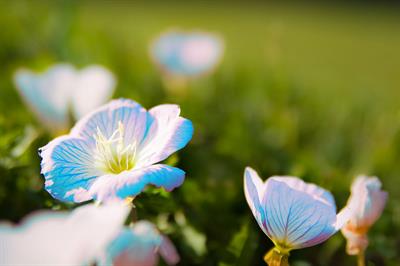}}&
  \imagetop{\includegraphics[width=.16\linewidth]{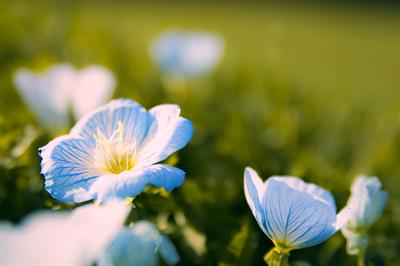}}&
  \imagetop{\includegraphics[width=.16\linewidth]{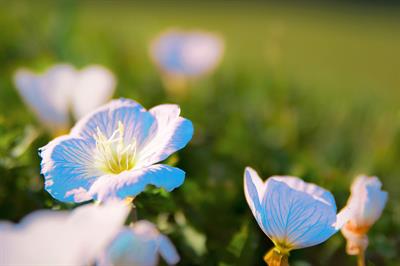}}&
  \imagetop{\includegraphics[width=.16\linewidth]{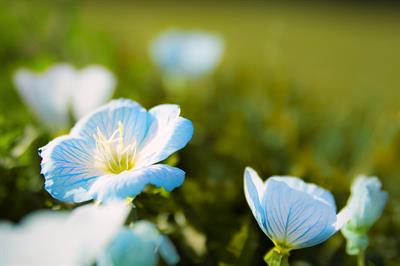}} \vspace{1pt}\\

  \imagetop{\includegraphics[width=.16\linewidth]{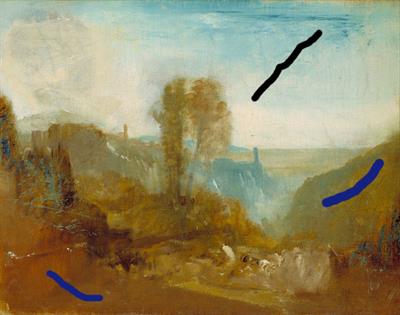}} 
  \imagetop{\includegraphics[width=.015\linewidth,frame]{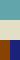}}&
  \imagetop{\includegraphics[width=.16\linewidth]{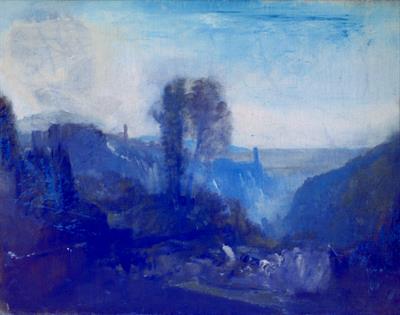}}&
  \imagetop{\includegraphics[width=.16\linewidth]{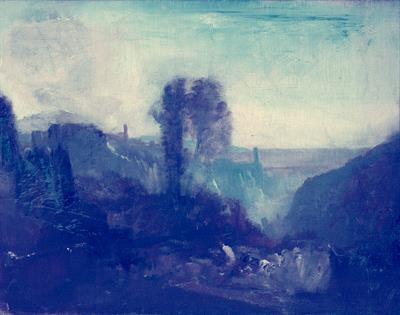}}&
  \imagetop{\includegraphics[width=.16\linewidth]{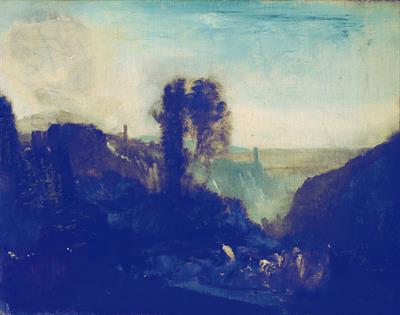}}&
  \imagetop{\includegraphics[width=.16\linewidth]{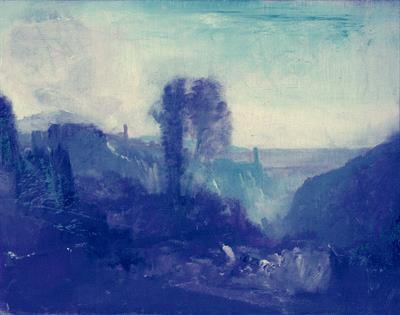}} \vspace{1pt}\\

  \imagetop{\includegraphics[width=.16\linewidth,height=0.12\linewidth]{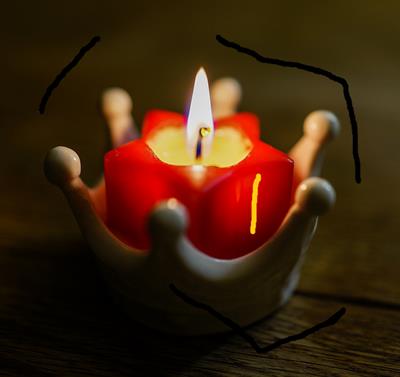}} 
  \imagetop{\includegraphics[width=.015\linewidth,frame]{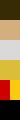}}&
  \imagetop{\includegraphics[width=.16\linewidth,height=0.12\linewidth]{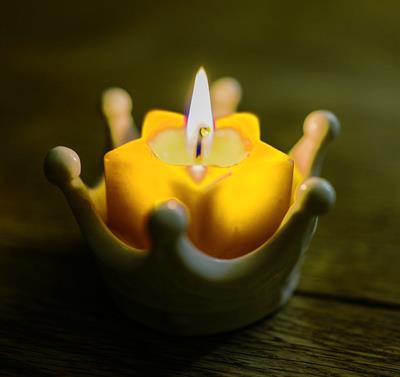}}&
  \imagetop{\includegraphics[width=.16\linewidth,height=0.12\linewidth]{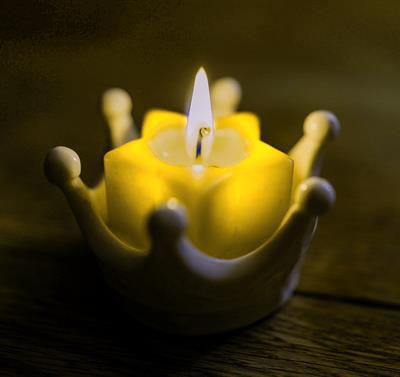}}&
  \imagetop{\includegraphics[width=.16\linewidth,height=0.12\linewidth]{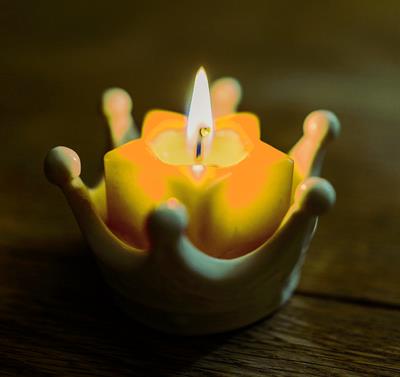}}&
  \imagetop{\includegraphics[width=.16\linewidth,height=0.12\linewidth]{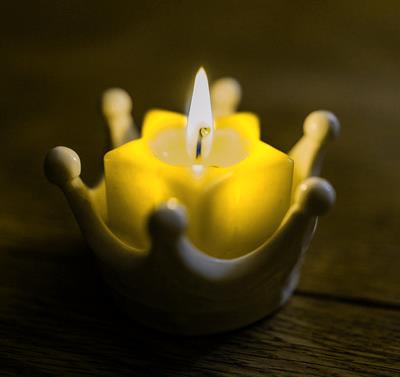}} \vspace{1pt}\\

  \end{tabular}
\caption{Image recoloring comparison. The first column shows the input strokes to ~\protect\cite{InstantProp} and the palette change for LayerBuilder, ~\protect\cite{MPEP}, and \protect\cite{PalettePhotoRecoloring}. Colored strokes indicate a change in color on the selected pixels, while black strokes constrain the pixels to stay the same color. Our method produces natural results even in the presence of challenging color blending scenarios. RBF interpolation~\protect\cite{InstantProp} and palette-based RBF interpolation~\protect\cite{PalettePhotoRecoloring} produces significant artifacts at color boundaries and blended regions.}
\label{fig:previousWorkComparison}
\vspace{-1.5em}
\end{figure*}

We first compare our image recoloring results with those of three previous algorithms: RBF interpolation~\cite{InstantProp}, palette-based RBF interpolation~\cite{PalettePhotoRecoloring}, and manifold-preserving edit propagation~\cite{MPEP}. RBF interpolation uses the common scribble-based paradigm for edit propagation, where a sparse set of colored strokes are propagated throughout the input. Palette-based RBF interpolation uses the same palette editing input as our method, and manifold-preserving edit propagation has been demonstrated on both types of inputs.  Because of the difference in inputs, it is difficult to make direct comparisons. We instead attempt to perform equivalent edits across the three methods; scribble input is provided to RBF interpolation while color palette edits are provided to LayerBuilder, manifold-preserving edit propagation, and palette-based RBF interpolation.

We ran each method on a variety of paintings and photographs. Figure~\ref{fig:previousWorkComparison} shows recoloring results from different methods on five images from this set. For more examples, please see the supplementary materials. We used the same parameters for all LayerBuilder results, except for the input palette. For the previous methods, we show the best result that does not overshoot the target region for recoloring, after varying parameters (such as the number of nearest neighbors $K$ for manifold-preserving edit propagation~\cite{MPEP} and color weight $\sigma_c$  for RBF interpolation~\cite{InstantProp}). We use the default grid dimension of 12 for palette-based RBF interpolation~\cite{PalettePhotoRecoloring}.

Our method preserves color transitions and blends as well as manifold-preserving edit propagation and is much faster. After layer computation, any color change only requires computing a weighted sum of layer colors for each pixel, while manifold-preserving edit propagation requires solving a large linear system. In addition, because LayerBuilder penalizes negative layer weights, layers can often be recolored to a dramatic degree, as seen in the soccer image. Manifold-preserving edit propagation does not prevent propagation of negative changes, so changing one color is more likely to lead to unintuitive changes elsewhere, like the greener grass in the soccer image. RBF interpolation and palette-based RBF interpolation work well on sharp regions and with small hue changes in blended regions. However, they are more prone to halo artifacts at color blending boundaries as seen in Figure~\ref{fig:previousWorkComparison}. Palette-based RBF interpolation does not consider spatial distance when shifting pixel colors, which results in a faster setup time than LayerBuilder, but can also lead to more cross-contamination when editing similarly colored objects (e.g. the red car and the maroon soccer jerseys in the soccer image).

\subsection{Comparisons with Matting and Illumination-Based Layers}

\begin{figure}[tb]
\ifdefined\showBigFigures
\vspace{-0.1in}
\renewcommand{\tabcolsep}{2pt}
\renewcommand{\arraystretch}{0.5}
 \begin{tabular}{c|ccc}
  Input & KNN Matting & LayerBuilder\\
  \imagetop{\includegraphics[width=.3\linewidth]{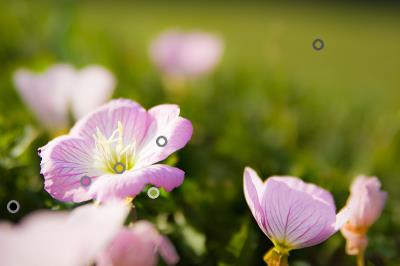}} 
  \imagetop{\includegraphics[width=.033\linewidth,frame]{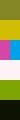}}&
  \imagetop{\includegraphics[width=.3\linewidth]{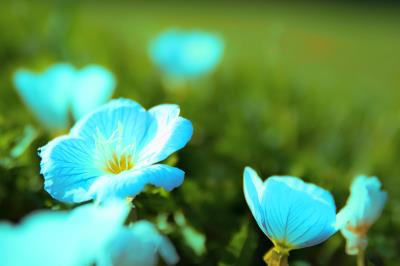}}&
  \imagetop{\includegraphics[width=.3\linewidth]{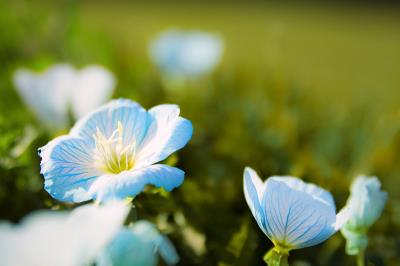}}\\
   \\ \hline
  \raisebox{-4\height}{Pink Layer}&
  \imagetop{\includegraphics[width=.3\linewidth,frame]{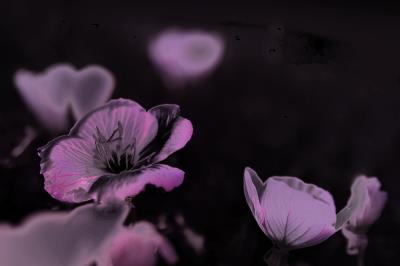}}&
  \imagetop{\includegraphics[width=.3\linewidth]{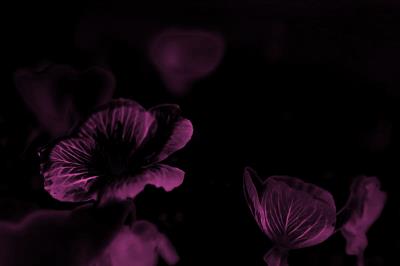}}\\
  \raisebox{-4\height}{Other Layers}&
  \imagetop{\includegraphics[width=.3\linewidth,frame]{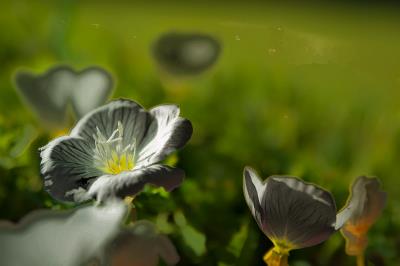}}&
  \imagetop{\includegraphics[width=.3\linewidth]{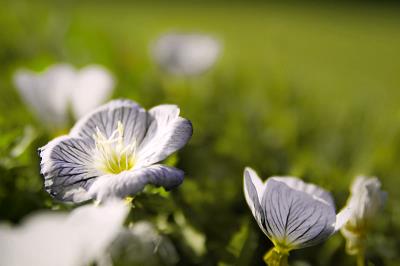}}
\end{tabular}
\fi
\caption{Image recoloring comparison between KNN matting~\protect\cite{KNNMatting} and LayerBuilder. The seed points used for KNN matting are shown along with our layer colors and color edits. Our layers better separate the pink pigment from the white, resulting in a more natural coloring.
\vspace{-1.5em}
}
\label{fig:knnMatting}
\end{figure}

A similar idea of layering is used in image matting to separate objects or materials in an image. Although these layers can be used for recoloring, they contain many different colors. We believe color-based layers are more often suited to recoloring tasks as they offer finer control over individual colors. In Figure~\ref{fig:knnMatting}, we compare recoloring using layers from LayerBuilder and KNN matting~\cite{KNNMatting}. The KNN pink layer mixes colors with the white flower, resulting in a more overblown recoloring. Our color-based layers can better separate the pink pigment from the rest of the flower and from the defocused areas.

\begin{figure}[tb]
\ifdefined\showBigFigures
\renewcommand{\tabcolsep}{2pt}
\renewcommand{\arraystretch}{0.5}
 \begin{tabular}{c|ccc}
  Input & ~\cite{Carroll2011illumination} & LayerBuilder\\
  \imagetop{\includegraphics[width=.3\linewidth]{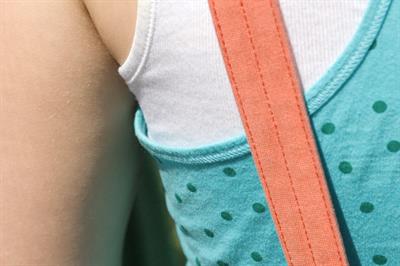}} 
  \imagetop{\includegraphics[width=.033\linewidth,frame]{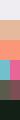}}&
  \imagetop{\includegraphics[width=.3\linewidth]{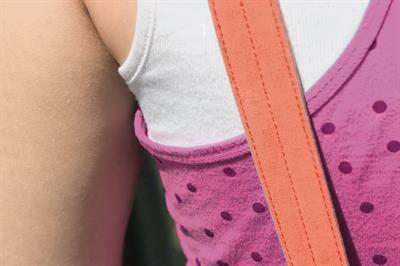}}&
  \imagetop{\includegraphics[width=.3\linewidth]{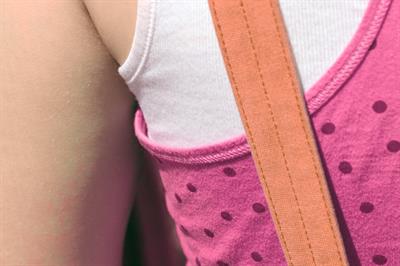}}\\
  \imagetop{\includegraphics[width=.3\linewidth]{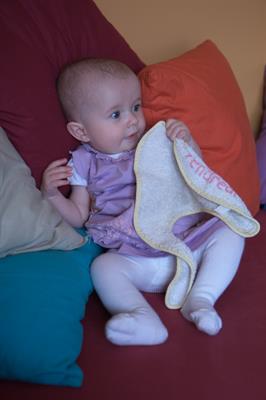}} 
  \imagetop{\includegraphics[width=.033\linewidth,frame]{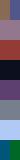}}&
  \imagetop{\includegraphics[width=.3\linewidth]{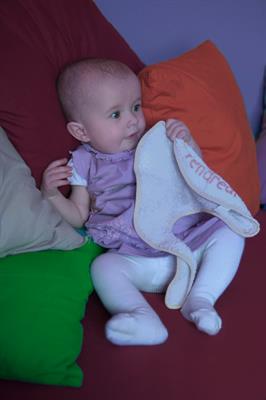}}&
  \imagetop{\includegraphics[width=.3\linewidth]{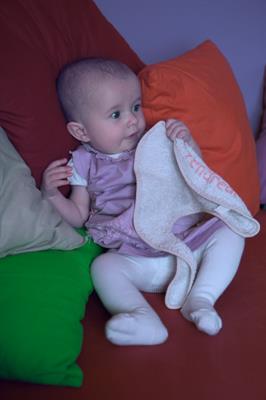}}
\end{tabular}
\fi
\caption{Image recoloring comparison between ~\protect\cite{Carroll2011illumination} and LayerBuilder. We show the original image and the palette edits used for our method in the first column. Our method can achieve plausible recolorings for pronounced interreflection effects while having a simpler editing pipeline. 
}
\vspace{-1em}
\label{fig:interreflections}
\end{figure}

Finally, we compare our method to the method by Carroll and colleagues~\shortcite{Carroll2011illumination}, which focuses on the problem of recoloring indirect reflection colors to be consistent with edited object colors and involves first decomposing the image into reflectance and illumination components. Figure~\ref{fig:interreflections} shows recoloring comparisons on two or their images, which feature color blending due to interreflections. Carroll and colleagues' method can better handle more subtle interreflections in the baby photo due to their separate handling of reflectance and illumination. However, we believe our method still achieves plausible results for more pronounced interreflection effects, such as in the clothes photo, as well as for other types of color blending without requiring a decomposition into reflectance and illumination images.

\subsection{Timing and Complexity}

\begin{table}[tb]
\centering
    \begin{tabular}{lc}
    \multicolumn{1}{c}{Stage}                         & Time  \\ \hline
    Superpixel computation        & 14.9s \\ 
    Per-pixel LLE weighting       & 7.2s  \\
    Linear solve (all iterations) & 8.4s  \\
    \end{tabular}
    \caption{Timings for different stages of our layer extraction algorithm.}
    \label{table:timing}
\end{table}

We evaluate our layer extraction algorithm on a PC with an Intel Core i7 3.7GHz processor and 8GB of RAM. We did not optimize or parallelize our layer extraction implementation. Timings for computing a five-layer decomposition of a 720x405, 70 frame video (20M pixels) are given in Table~\ref{table:timing}. We use a single-threaded version of Eigen for our linear solve\footnote{eigen.tuxfamily.org/}. Because our solve is performed on superpixels, the size of the linear system remains manageable (20000 free variables).  Although the time required to compute the per-pixel linear-embedding weights is significant, it's worth noting that these are computed independently for each pixel and could easily be parallelized. Our peak memory usage was less than 700MB, much of which is needed to store the layers for each video frame without compression. Once the layers are computed, color edits can be computed at interactive rates (less than 5ms per frame).

\subsection{Varying Superpixel Count}
\label{sec:superpixelCount}

\begin{figure*}[tb]

\ifdefined\showBigFigures
\renewcommand{\arraystretch}{0.5}
  \begin{tabular}{c|ccc}
  \centering
  \multirow{2}{*}{\raisebox{-1.05\height}{\includegraphics[width=.27\linewidth]{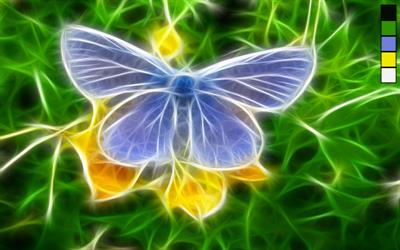}}} &
  \imagetop{\includegraphics[width=.21\linewidth]{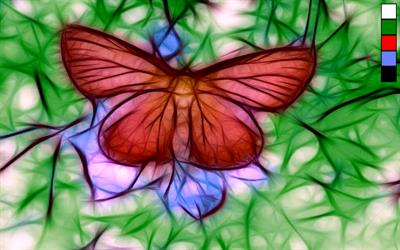}}&
	\imagetop{\includegraphics[width=.21\linewidth]{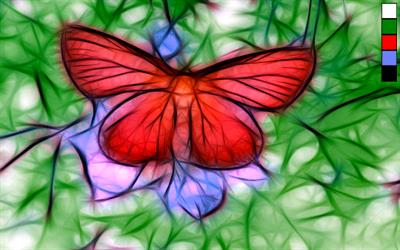}}&
	\imagetop{\includegraphics[width=.21\linewidth]{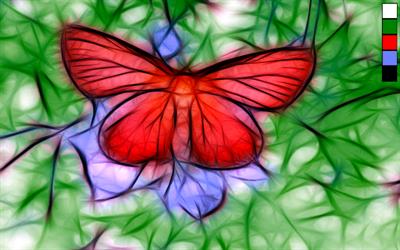}}\\
  &
  \imagetop{\includegraphics[width=.21\linewidth]{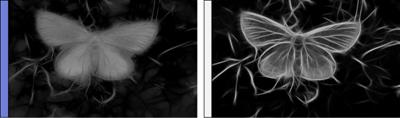}}&
	\imagetop{\includegraphics[width=.21\linewidth]{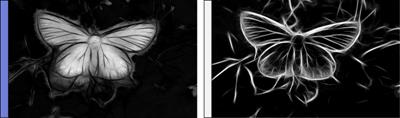}}&
	\imagetop{\includegraphics[width=.21\linewidth]{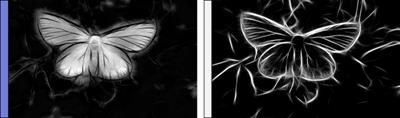}}\\ \\
  Original Image & 50 Superpixels & 250 Superpixels & 2000 Superpixels \\
  \end{tabular}
\fi
\caption{Changing the number of superpixels. Top: recoloring result with each superpixel count. Bottom: two of the extracted layers, corresponding to the blue and white colors in the original image. Using too few superpixels can result in layers that are not well separated.}
\label{fig:superpixelCount}
\vspace{-1.5em}
\end{figure*}

In Figure~\ref{fig:superpixelCount}, we investigate the effect of superpixel count on the extracted layers and the quality of the recolored image. With 50 superpixels, the white and blue layers are not correctly separated and overlap significantly. Recoloring with these layers results in undesirable color blending and does not produce the expected saturated shade of red requested in the recolored palette. As the number of superpixels increases, the layers separate more cleanly and this improvement is reflected in the recolored image. For images shown in the paper, the extracted layers do not change significantly when more than 2000 superpixels are used.

%% file: applications.tex
\section{Applications}

We demonstrate our layer decomposition algorithm on more complex recoloring of images and video, and give examples for three other applications - automated pattern coloring, texture synthesis, and layer filtering.

\begin{figure*}[htb!]
\ifdefined\showBigFigures
\vspace{-0.1in}
\hspace{-0.5in}
\renewcommand{\tabcolsep}{5.5pt}
  \begin{tabular}{c|c}
    \begin{tabular}{cc|cc}
    \raisebox{0.1\height}{\rot{{\small Frame 2}}} & \includegraphics[width=.13\linewidth]{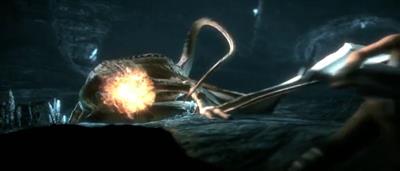} & \includegraphics[width=.13\linewidth]{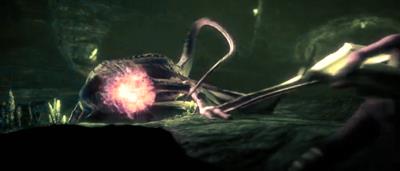} & \includegraphics[width=.13\linewidth]{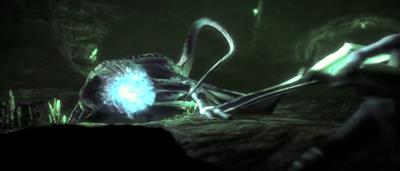} \\
    \raisebox{0.02\height}{\rot{{\small Frame 12}}} & \includegraphics[width=.13\linewidth]{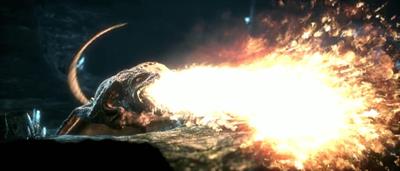} & \includegraphics[width=.13\linewidth]{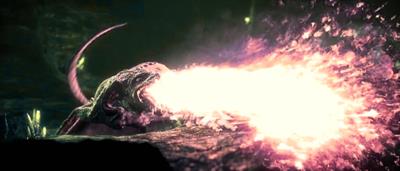} & \includegraphics[width=.13\linewidth]{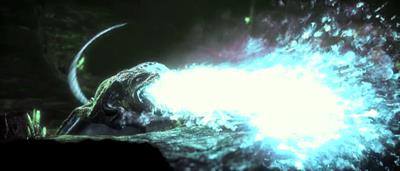} \\
    \raisebox{0.02\height}{\rot{{\small Frame 22}}} & \includegraphics[width=.13\linewidth]{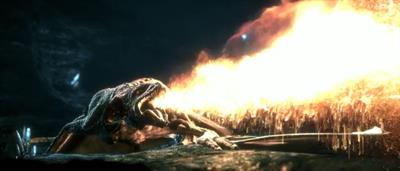} & \includegraphics[width=.13\linewidth]{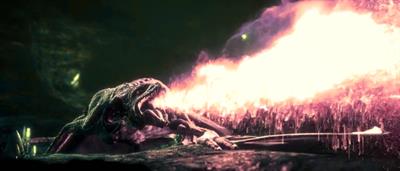} & \includegraphics[width=.13\linewidth]{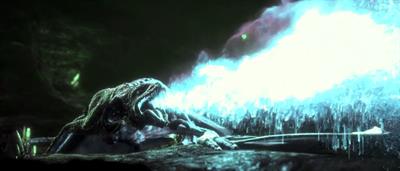} \\
    & \includegraphics[width=.13\linewidth]{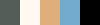} & \includegraphics[width=.13\linewidth]{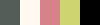} & \includegraphics[width=.13\linewidth]{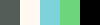} \\
    \end{tabular}
    
    & 

    \begin{tabular}{cc|cc}
    \raisebox{0.2\height}{\rot{{\small Frame 53}}}  & \includegraphics[width=.13\linewidth]{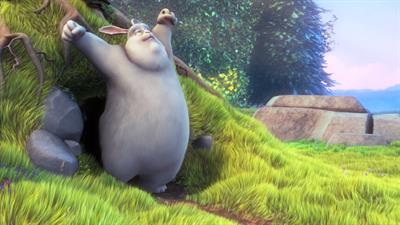} & \includegraphics[width=.13\linewidth]{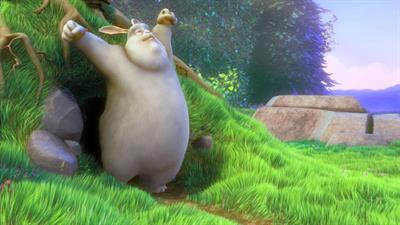} & \includegraphics[width=.13\linewidth]{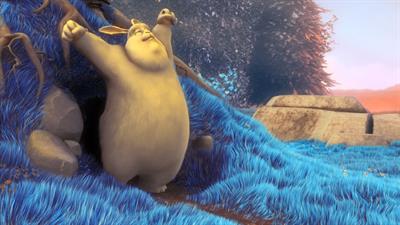}\\
    \raisebox{0.1\height}{\rot{{\small Frame 110}}}  & \includegraphics[width=.13\linewidth]{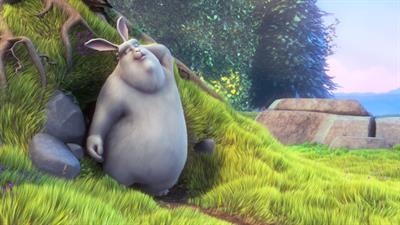} & \includegraphics[width=.13\linewidth]{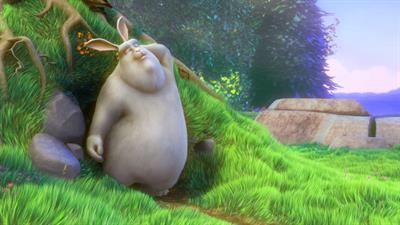} & \includegraphics[width=.13\linewidth]{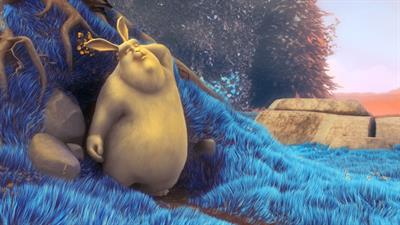}\\
    \raisebox{0.12\height}{\rot{{\small Frame 197}}}  & \includegraphics[width=.13\linewidth]{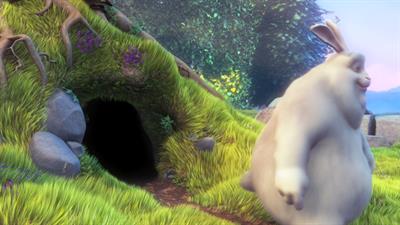} & \includegraphics[width=.13\linewidth]{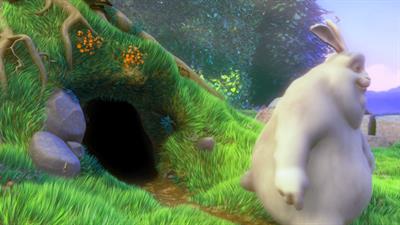} & \includegraphics[width=.13\linewidth]{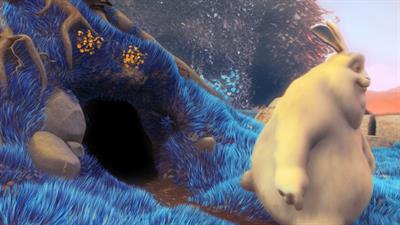}\\
    & \includegraphics[width=.13\linewidth]{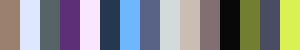} & \includegraphics[width=.13\linewidth]{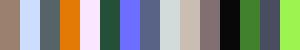} & \includegraphics[width=.13\linewidth]{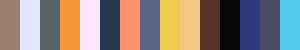} \\
    \end{tabular}
  \end{tabular}
\fi
\caption{Video recoloring. Left Columns: frames from the original video. Middle and Right Columns: frames from edited videos. Changes in the color palettes of the videos, visualized below, propagate through the entire video in real-time using the layer decomposition.}
\label{fig:video}
\end{figure*}

\subsection{Interactive Recoloring with Layers}

It is straightforward to apply our layer decomposition algorithm to interactively recolor images and videos. Once the layering is known, recoloring an image or video frame with a new palette is as simple as performing a linear combination of $\layerCount$ images. Figure~\ref{fig:teaser} shows two image recoloring examples for natural and digital images. The palette and layers were automatically computed with no additional user constraints. The algorithm produces pleasing results, even when undergoing significant color variation such as changing the season in a photograph from summer to fall.

In Figure~\ref{fig:video}, we show two examples of video recoloring. Our method supports significant edits on large palettes, such as modifying the grass from green to blue and adapting the other colors in a pleasing fashion. As demonstrated in the accompanying video, the real-time feedback to changes made to the palette helps users generate visually appealing results when editing both images and video. Even though we use only 4000 superpixels in all these images and videos, the resulting images do not exhibit significant spatial or temporal artifacts from this superpixel representation.

\subsection{Pattern Recoloring}
\label{sec:patternRecoloring}

\begin{figure*}[htb!]
  \ifdefined\showBigFigures
    \begin{tabular}{c|cccccc}
      \includegraphics[width=0.12\linewidth]{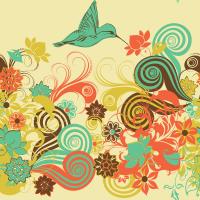}&
      \includegraphics[width=0.12\linewidth]{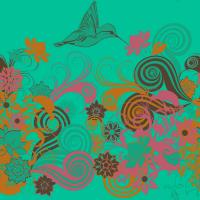}&
      \includegraphics[width=0.12\linewidth]{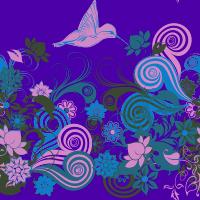}&
      \includegraphics[width=0.12\linewidth]{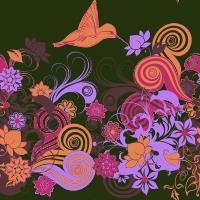}&
      \includegraphics[width=0.12\linewidth]{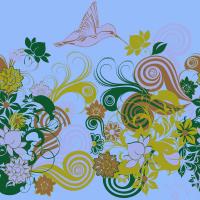}&
      \includegraphics[width=0.12\linewidth]{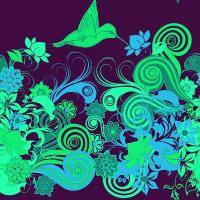}&
      \includegraphics[width=0.12\linewidth]{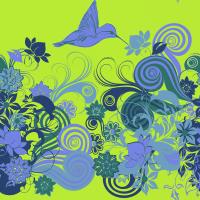}\\

      \includegraphics[width=0.12\linewidth,height=0.7in]{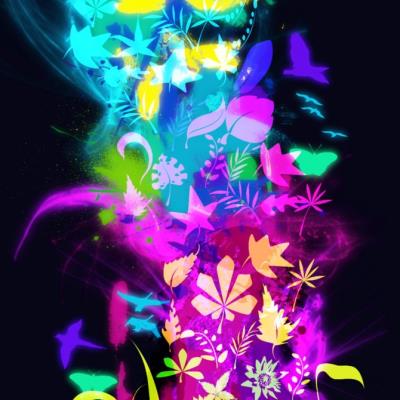}&
      \includegraphics[width=0.12\linewidth,height=0.7in]{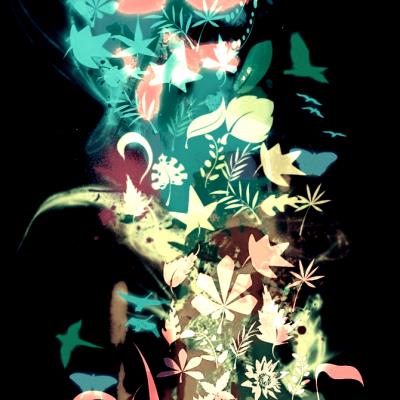}&
      \includegraphics[width=0.12\linewidth,height=0.7in]{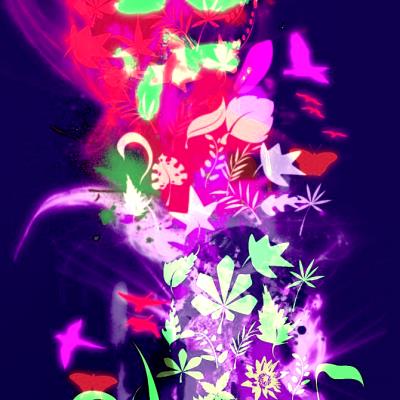}&
      \includegraphics[width=0.12\linewidth,height=0.7in]{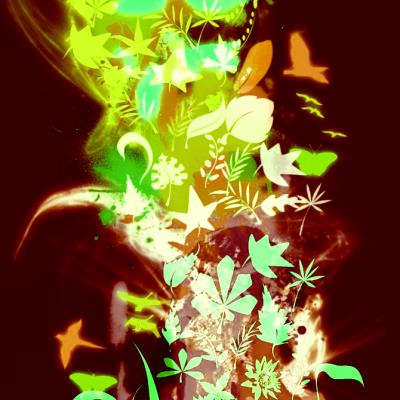}&
      \includegraphics[width=0.12\linewidth,height=0.7in]{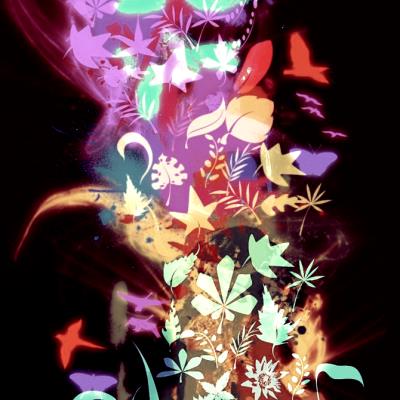}&
      \includegraphics[width=0.12\linewidth,height=0.7in]{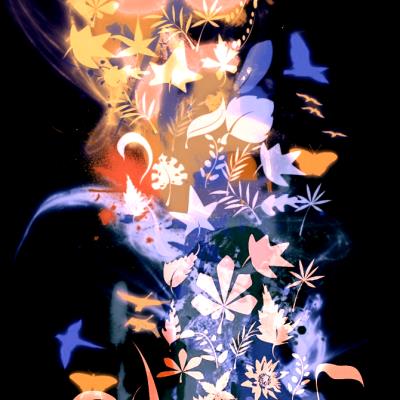}&
      \includegraphics[width=0.12\linewidth,height=0.7in]{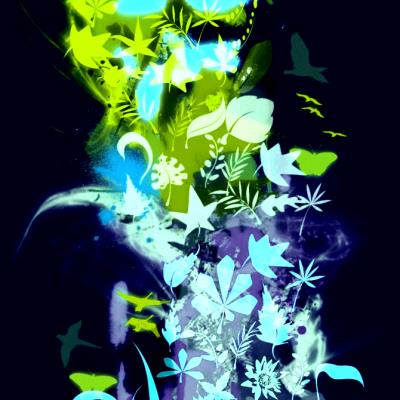}\\

      \includegraphics[width=0.12\linewidth]{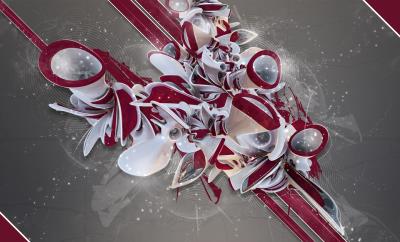}&
      \includegraphics[width=0.12\linewidth]{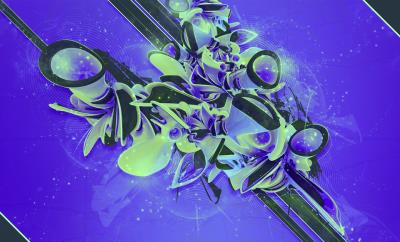}&
      \includegraphics[width=0.12\linewidth]{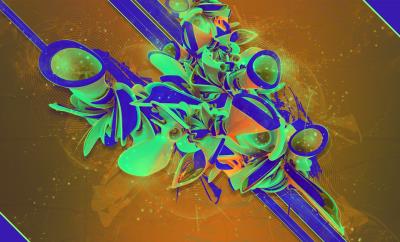}&
      \includegraphics[width=0.12\linewidth]{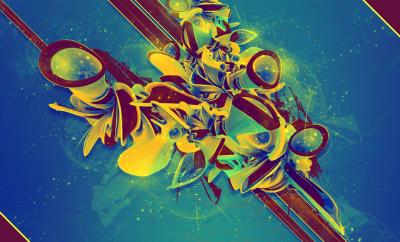}&
      \includegraphics[width=0.12\linewidth]{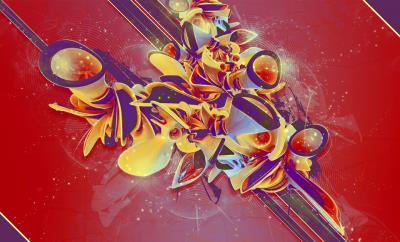}&
      \includegraphics[width=0.12\linewidth]{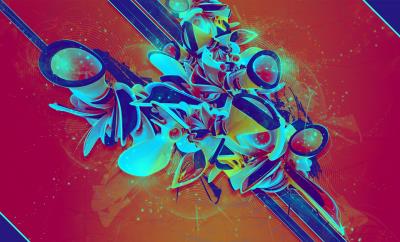}&
      \includegraphics[width=0.12\linewidth]{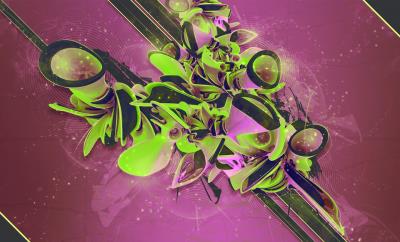}\\

      \includegraphics[width=0.12\linewidth]{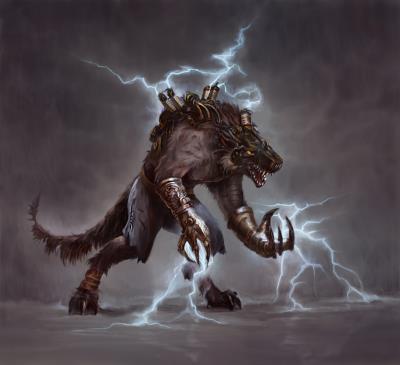}&
      \includegraphics[width=0.12\linewidth]{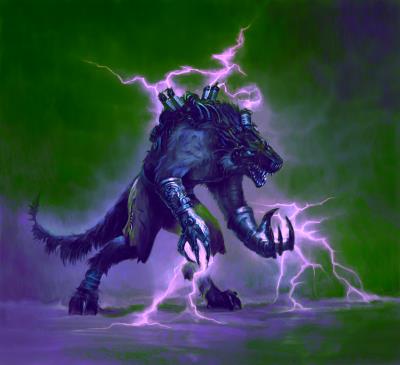}&
      \includegraphics[width=0.12\linewidth]{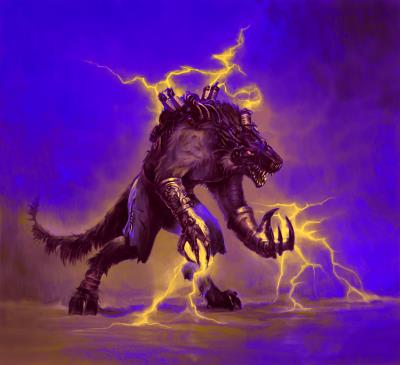}&
      \includegraphics[width=0.12\linewidth]{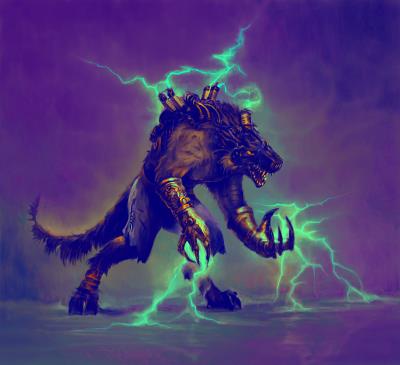}&
      \includegraphics[width=0.12\linewidth]{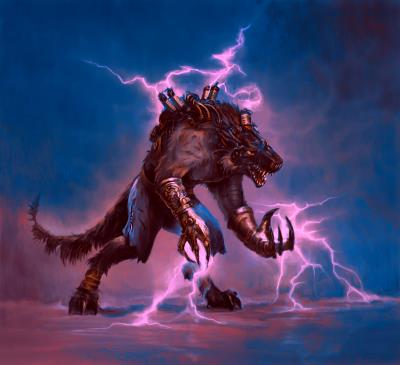}&
      \includegraphics[width=0.12\linewidth]{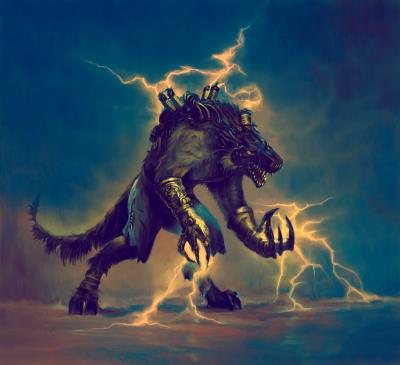}&
      \includegraphics[width=0.12\linewidth]{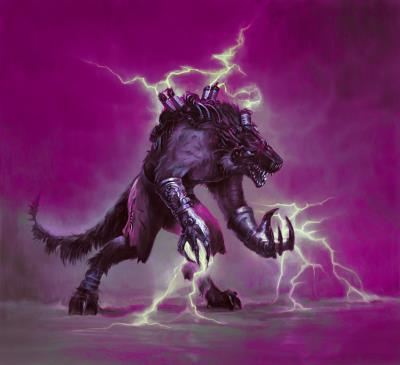}\\

    \end{tabular}
  \fi
	\caption{Recoloring suggestions from model trained on 907 COLOURLovers patterns, using layer properties as features. Except for the first example, luminance was constrained to be close to the original. We use an automatic palette.}
	\label{fig:designExploration}
\end{figure*}

In addition to allowing users to interactively recolor an image, we can also automatically generate coloring suggestions for that image to help users better explore the coloring state space. Lin and colleagues propose a graphical model for scoring pattern colorings, where images are represented as non-overlapping regions of solid color~\shortcite{PatternColoring}. The model scores colorings by learning color distributions over individual and adjacent regions from a training dataset of images, where each region is represented as a set of shape features. However, it can only recolor images that are accompanied by a layer decomposition which is not available for most images. We adapt this model to support arbitrary images by adding features that take into account overlapping image layers from LayerBuilder. See the supplemental materials for the list of features considered. Figure~\ref{fig:designExploration} shows coloring suggestions for four images sampled from the adapted model which was trained on 907 COLOURLovers patterns. By using our layer decomposition algorithm, we can automatically recolor a much wider variety of images.

\subsection{Texture Synthesis}

\begin{figure}[htb!]
  \begin{center}
    \ifdefined\showBigFigures
      \begin{tabular}{c|cc}
        Exemplar & RGB & RGB \& Layers \\
        \hline \hline \\
        \includegraphics[width=.17\linewidth]{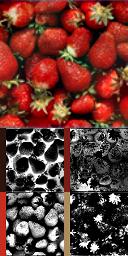} & \includegraphics[width=.34\linewidth]{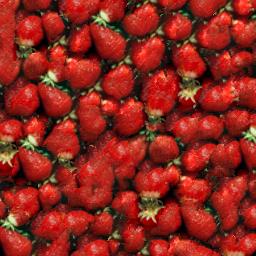} 
        & \includegraphics[width=.34\linewidth]{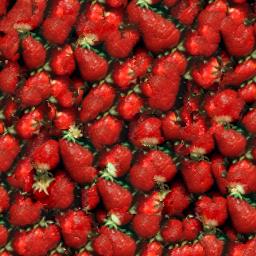} \\

        \includegraphics[width=.17\linewidth]{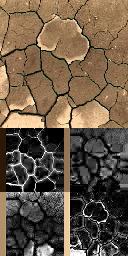} & \includegraphics[width=.34\linewidth]{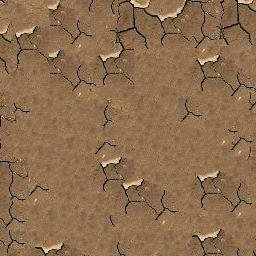} 
        & \includegraphics[width=.34\linewidth]{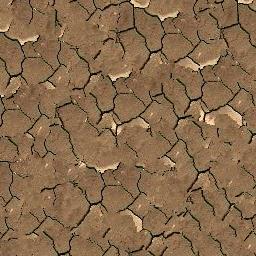} \\

        \includegraphics[width=.17\linewidth]{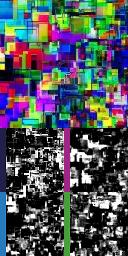} & \includegraphics[width=.34\linewidth]{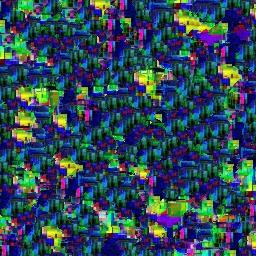} 
        & \includegraphics[width=.34\linewidth]{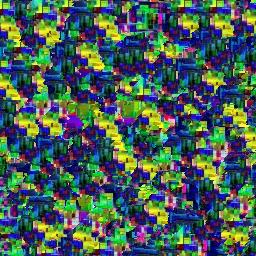} \\

      \end{tabular}
    \fi
  \end{center}
  \caption{Texture synthesis augmented with layers. Left: exemplar textures and their layer decompositions. An automatic palette was used. Middle: results of texture synthesis using pixel RGB values. Right: synthesis using RGB and the layer contribution vector from each pixel.}
  \label{fig:texsyn}
  \vspace{-1.5em}
\end{figure}

Texture synthesis algorithms traditionally synthesize an image by comparing incomplete pixel neighborhoods in the result with existing neighborhoods in an exemplar. These neighborhoods are often computed in RGB space, but can easily be extended to incorporate image layer features to provide a more informative representation of texture features. We adapt the approach proposed by Lefebvre and Hoppe ~\shortcite{TextureSynthesis} to compute feature vectors for each neighborhood, including both the RGB values at each pixel and the layer weights corresponding to each palette color. Since small spatial neighborhoods cannot encode large texture features or semantic structure, appearance-space texture synthesis and other approaches use the notion of a user-provided feature mask to guide synthesis. Our layer decompositions automatically generate layers that encode the same types of detail represented by these feature masks. Figure~\ref{fig:texsyn} shows appearance-space texture synthesis results using only RGB features and using both RGB and layer features. Although synthesis remains challenging for images with non-stationary features, for images such as the ground pattern (Figure~\ref{fig:texsyn} second row) incorporating the layers into the local feature vector better captures the distribution implied by the exemplar. 

\subsection{Layer Filtering}

\begin{figure}[htb!]
\ifdefined\showBigFigures
	\begin{center}
	\begin{tabular}{c|c|c}
	\includegraphics[width=.3\linewidth]{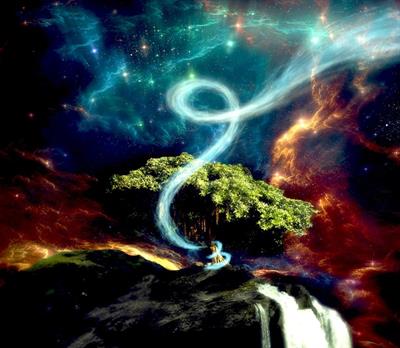} & \includegraphics[width=.3\linewidth]{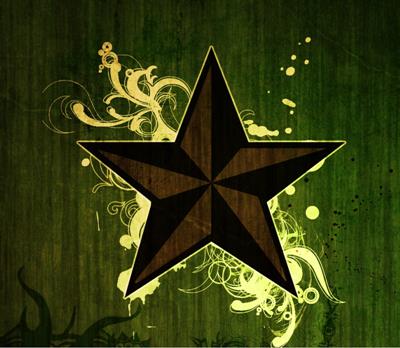} & \includegraphics[width=.3\linewidth]{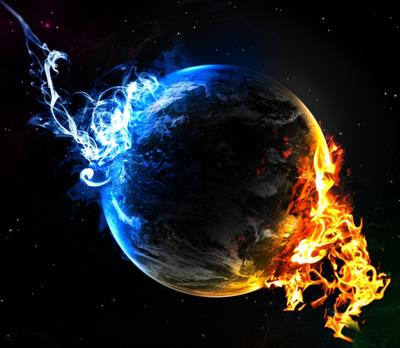} \\

	\includegraphics[width=.3\linewidth]{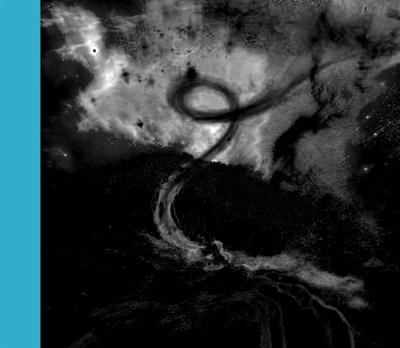} & \includegraphics[width=.3\linewidth]{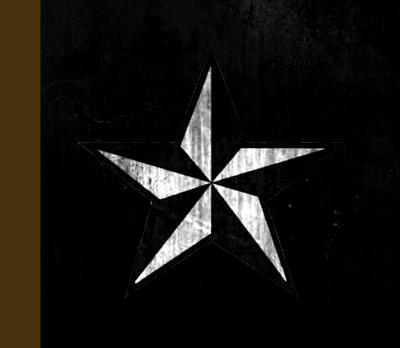} & \includegraphics[width=.3\linewidth]{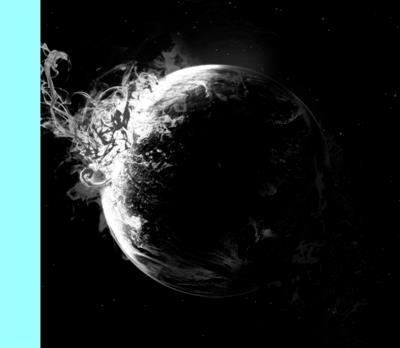} \\

	$\vert$ & $\vert$ & $\vert$ \\
	Gaussian Filter & Emboss Filter & Motion Blur \\
	$\downarrow$ & $\downarrow$ & $\downarrow$ \\

	\includegraphics[width=.3\linewidth]{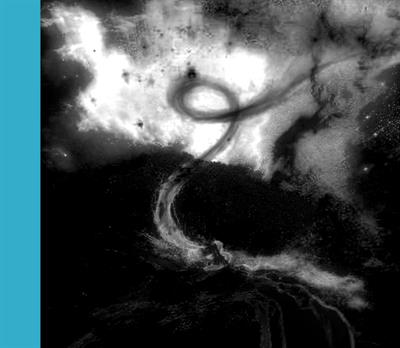} & \includegraphics[width=.3\linewidth]{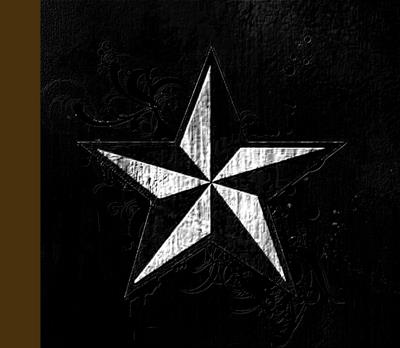} & \includegraphics[width=.3\linewidth]{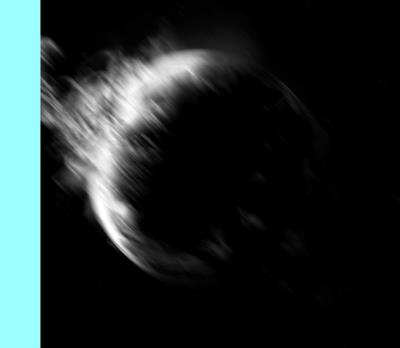} \\

	\includegraphics[width=.3\linewidth]{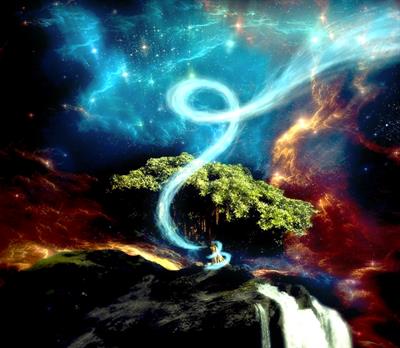} & \includegraphics[width=.3\linewidth]{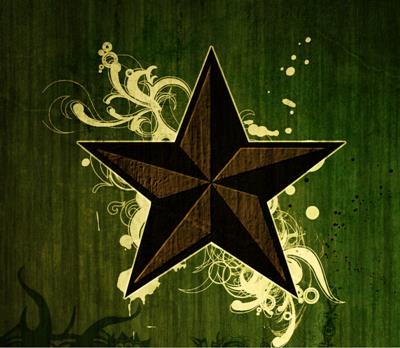} & \includegraphics[width=.3\linewidth]{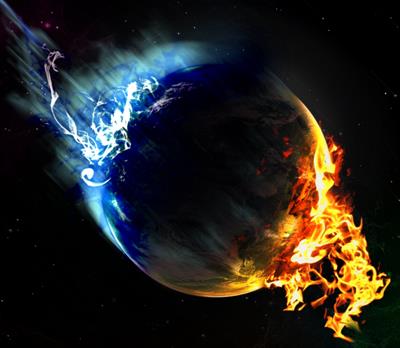} \\
	\end{tabular}
	\end{center}
\fi
	\caption{Filtering layers. Individual layers can be filtered, and the image reconstructed with the filtered layers. In this case, a single layer from each image is filtered (rows 2 to 3), and the resulting reconstruction is shown in row 4.}
	\label{fig:layerfiltering}
  \vspace{-1.5em}
\end{figure}

Various image editing operations can be performed in a localized manner by editing individual layers, and then recomposing them. In Figure~\ref{fig:layerfiltering}, we show the effects of several filtering operations on individual layers: a Gaussian filter emphasizes the light in the sky, the star is embossed, and the blue fire is motion blurred, respectively. Although the edited layers are tightly blended into the image, our layers correctly separate the different color sources in the image and these high-level operations produce few unintended artifacts.

%% file: discussion.tex
\begin{figure}[tb]

\ifdefined\showBigFigures
	\includegraphics[width=\linewidth]{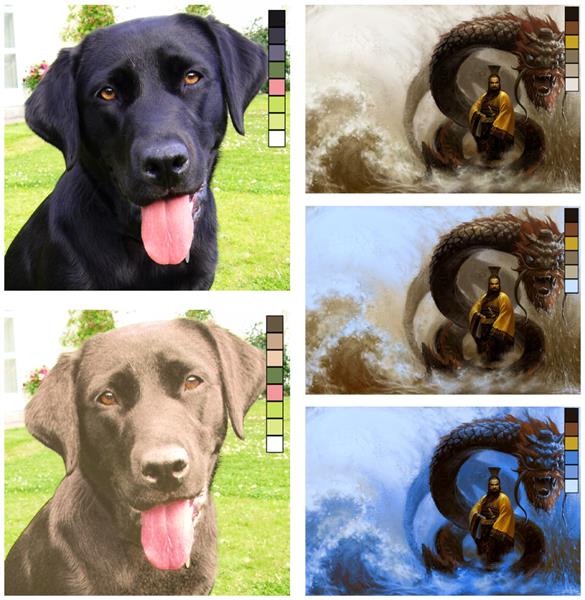}
\fi
\caption{Limitations of our method. The top row shows the original images. \emph{Left column:} an attempt to recolor a black Labrador to a chocolate Labrador. \emph{Right column:} two attempts to turn the sky and water from brown to blue.}
\label{fig:failure}
\vspace{-2em}
\end{figure}

\section{Limitations}
\label{sec:failure}

Although the manifold-preserving constraint we use is effective at handling color blending and object motion scenarios, our algorithm can have difficulty automatically decomposing and recoloring several types of photographs or digital images. In Figure~\ref{fig:failure}, we show two recoloring failures where no user constraints were used to separate the layers. When the layers represent complex lighting such as specular reflections on fur, attempting to recolor the layers can fail to produce a pleasing image. When the colors represent semantically distinct but similarly colored regions, such as a brown dragon on a brown background, it can be difficult to change the background without significant effects on the foreground without using user constraints, as shown in Figure~\ref{fig:constraints}.

User constraints can be used to help separate similarly colored regions, and they work best when the regions are relatively large and spatially distant. Due to the granularity of the superpixel manifold, our approach can have trouble separating similarly colored regions that small (e.g. only composed of a few superpixels) and are close together. In these cases, a user-provided mask can be used to separate regions. Object-based layers from image matting can also be combined with our layers to better localize edits.

\section{Discussion and Future Work}

In this paper, we present a method for interactive color editing of images and videos by using locally linear embedding to decompose the input into colored layers. By preserving the overall pixel manifold structure instead of local pixel affinities, we produce layers that can handle significant color blending. We demonstrate the usefulness of these layers for interactive recoloring, suggesting automatic recolorings, texture synthesis, and filter-based editing.

We compute a set of additive colored layers whose linear combination reconstructs the original image. While physical lighting is well-modeled by an additive process, many digital layers are composited using non-linear effects. Instead of recovering an order-independent set of linearly combined layers, in future work we look to explore ways to simultaneously extract both the layers and their order-dependent compositing function. We believe that such a formulation has the potential to significantly improve the process of editing images and videos.

\section{Acknowledgements}
Support for this research was provided by Intel (ISTC-VC). Input images are by Flickr users Guy Schmickle, tambako, alohateam, and nanagyei; Noah Bradley; DeviantArt users Elektroll, NeocoreGames, and coreyDA; COLOURLovers user symea; and the Blender Foundation.

%% file: layer-extraction-arxiv.bbl
\newcommand{\etalchar}[1]{$^{#1}$}
\begin{thebibliography}{\uppercase{RAGS01}}

\bibitem[AB94]{SRG}
\textsc{Adams R., Bischof L.}:
\newblock Seeded region growing.
\newblock \emph{Pattern Analysis and Machine Intelligence, IEEE Transactions on
  16}, 6 (1994), 641--647.

\bibitem[Adl88]{boxConstraints}
\textsc{Adlers M.}:
\newblock \emph{Sparse Least Squares Problems with Box Constraints}.
\newblock Linkopings Universitet, 1988.

\bibitem[AP08]{AppProp}
\textsc{An X., Pellacini F.}:
\newblock Appprop: all-pairs appearance-space edit propagation.
\newblock In \emph{ACM Transactions on Graphics (TOG)} (2008), vol.~27, ACM,
  p.~40.

\bibitem[BPD09]{Bousseau2009User}
\textsc{Bousseau A., Paris S., Durand F.}:
\newblock User-assisted intrinsic images.
\newblock In \emph{ACM Transactions on Graphics (TOG)} (2009), vol.~28, ACM,
  p.~130.

\bibitem[BV04]{convexOptimization}
\textsc{Boyd S.~P., Vandenberghe L.}:
\newblock \emph{Convex optimization}.
\newblock Cambridge university press, 2004.

\bibitem[CCSS01]{BayesianMatting}
\textsc{Chuang Y.-Y., Curless B., Salesin D.~H., Szeliski R.}:
\newblock A {B}ayesian approach to digital matting.
\newblock In \emph{In Proc. Computer Vision and Pattern Recognition} (2001),
  vol.~2, IEEE, pp.~II--264.

\bibitem[CFL{\etalchar{*}}15]{PalettePhotoRecoloring}
\textsc{Chang H., Fried O., Liu Y., DiVerdi S., Finkelstein A.}:
\newblock Palette-based photo recoloring.
\newblock \emph{ACM Transactions on Graphics (Proc. SIGGRAPH) 34}, 4 (July
  2015).

\bibitem[CLT13]{KNNMatting}
\textsc{Chen Q., Li D., Tang C.-K.}:
\newblock {KNN} matting.
\newblock \emph{Pattern Analysis and Machine Intelligence, IEEE Transactions on
  35}, 9 (2013), 2175--2188.

\bibitem[CRA11]{Carroll2011illumination}
\textsc{Carroll R., Ramamoorthi R., Agrawala M.}:
\newblock Illumination decomposition for material recoloring with consistent
  interreflections.
\newblock In \emph{ACM Transactions on Graphics (TOG)} (2011), vol.~30, ACM,
  p.~43.

\bibitem[CZL{\etalchar{*}}14]{MPEPSparse}
\textsc{Chen X., Zou D., Li J., Cao X., Zhao Q., Zhang H.}:
\newblock Sparse dictionary learning for edit propagation of high-resolution
  images.
\newblock In \emph{Proceedings of the IEEE Conference on Computer Vision and
  Pattern Recognition} (2014), pp.~2854--2861.

\bibitem[CZZT12]{MPEP}
\textsc{Chen X., Zou D., Zhao Q., Tan P.}:
\newblock Manifold preserving edit propagation.
\newblock \emph{ACM Transactions on Graphics (TOG) 31}, 6 (2012), 132.

\bibitem[FFL10]{DiffusionMaps}
\textsc{Farbman Z., Fattal R., Lischinski D.}:
\newblock Diffusion maps for edge-aware image editing.
\newblock In \emph{ACM SIGGRAPH Asia 2010 Papers} (2010), SIGGRAPH ASIA '10,
  ACM.

\bibitem[LAA08]{ScribbleBoost}
\textsc{Li Y., Adelson E., Agarwala A.}:
\newblock Scribbleboost: Adding classification to edge-aware interpolation of
  local image and video adjustments.
\newblock In \emph{In Proc. Eurographics Symposium on Rendering} (2008),
  Eurographics Association, pp.~1255--1264.

\bibitem[LH06]{TextureSynthesis}
\textsc{Lefebvre S., Hoppe H.}:
\newblock Appearance-space texture synthesis.
\newblock In \emph{ACM Transactions on Graphics (TOG)} (2006), vol.~25, ACM,
  pp.~541--548.

\bibitem[LH13]{SharonPaletteExtraction}
\textsc{Lin S., Hanrahan P.}:
\newblock Modeling how people extract color themes from images.
\newblock CHI 2013, ACM.

\bibitem[LJH10]{InstantProp}
\textsc{Li Y., Ju T., Hu S.-M.}:
\newblock Instant propagation of sparse edits on images and videos.
\newblock In \emph{Computer Graphics Forum} (2010), vol.~29, pp.~2049--2054.

\bibitem[LL11]{MaterialMatting}
\textsc{Lepage D., Lawrence J.}:
\newblock Material matting.
\newblock In \emph{ACM Transactions on Graphics} (2011), vol.~30, ACM.

\bibitem[LLW04]{ScribbleColorization}
\textsc{Levin A., Lischinski D., Weiss Y.}:
\newblock Colorization using optimization.
\newblock In \emph{Proc. SIGGRAPH 2004} (2004).

\bibitem[LLW08]{ClosedFormImageMatting}
\textsc{Levin A., Lischinski D., Weiss Y.}:
\newblock A closed-form solution to natural image matting.
\newblock \emph{Pattern Analysis and Machine Intelligence, IEEE Transactions on
  30}, 2 (2008), 228--242.

\bibitem[LRFH13]{PatternColoring}
\textsc{Lin S., Ritchie D., Fisher M., Hanrahan P.}:
\newblock Probabilistic color-by-numbers: Suggesting pattern colorizations
  using factor graphs.
\newblock In \emph{ACM SIGGRAPH 2013 papers} (2013), SIGGRAPH '13.

\bibitem[RAGS01]{Reinhard2001}
\textsc{Reinhard E., Ashikhmin M., Gooch B., Shirley P.}:
\newblock Color transfer between images.
\newblock \emph{IEEE Comput. Graph. Appl. 21}, 5 (Sept. 2001), 34--41.

\bibitem[RS00]{LLE}
\textsc{Roweis S.~T., Saul L.~K.}:
\newblock Nonlinear dimensionality reduction by locally linear embedding.
\newblock \emph{Science 290}, 5500 (2000), 2323--2326.

\bibitem[SV11]{MultipleImageLayers}
\textsc{Singaraju D., Vidal R.}:
\newblock Estimation of alpha mattes for multiple image layers.
\newblock \emph{Pattern Analysis and Machine Intelligence, IEEE Transactions on
  33}, 7 (2011), 1295--1309.

\bibitem[TJT05]{Tai2005local}
\textsc{Tai Y.-W., Jia J., Tang C.-K.}:
\newblock Local color transfer via probabilistic segmentation by
  expectation-maximization.
\newblock In \emph{2005 IEEE Computer Society Conference on Computer Vision and
  Pattern Recognition (CVPR'05)} (2005), vol.~1, IEEE, pp.~747--754.

\bibitem[TLG16]{TAN2016DIL}
\textsc{Tan J., Lien J.-M., Gingold Y.}:
\newblock Decomposing images into layers via {RGB}-space geometry.
\newblock \emph{ACM Transactions on Graphics (TOG) 36}, 1 (Nov. 2016),
  7:1--7:14.

\bibitem[XLJ{\etalchar{*}}09]{KDTreeProp}
\textsc{Xu K., Li Y., Ju T., Hu S.-M., Liu T.-Q.}:
\newblock Efficient affinity-based edit propagation using {KD} tree.
\newblock In \emph{ACM Transactions on Graphics (TOG)} (2009), vol.~28, ACM,
  p.~118.

\end{thebibliography}
